\documentclass[10pt,aps,longbibliography,prd,superscriptaddress,twocolumn,amsmath,amssymb,floatfix,
nofootinbib 
]{revtex4-2}   
           
\usepackage{bm}
\usepackage{float}
\usepackage{enumerate}
\usepackage{subfigure}
\usepackage{color}
\usepackage{graphicx} 
\usepackage[dvipsnames]{xcolor}   
\newcommand{\la}{{\langle}}
\newcommand{\ra}{{\rangle}}
\usepackage{tcolorbox}
\usepackage{bbold}

\newcommand{\abs}[1]{|{#1}|} 
\usepackage{multibib}
\newcites{supp}{Supplementary References}
\usepackage[normalem]{ulem}
\usepackage{xcolor}  
\newcommand{\hlm}[1]{\textcolor{magenta}{#1}}     
 
\usepackage[colorlinks,citecolor=blue,linkcolor=blue,urlcolor=blue,plainpages=false,pdfpagelabels]{hyperref} 
\usepackage{orcidlink}
\begin{document} 
\title{Probing imbalanced Weyl nodes in a two-dimensional anisotropic semimetal via optical conductivity} 

\author{Suheel Ahmad Malik \orcidlink{0009-0000-1618-6602}}
\affiliation{Department of Physics, Jamia Millia Islamia, New Delhi-110025, INDIA}
\author{M.A.H Ahsan \orcidlink{0000-0002-9870-2769}}
\affiliation{Department of Physics, Jamia Millia Islamia, New Delhi-110025, INDIA}
\author{SK Firoz Islam \orcidlink{0000-0003-1224-622X}}
\affiliation{Department of Physics, Jamia Millia Islamia, New Delhi-110025, INDIA}
\date{\today}
\begin{abstract}
We present a theoretical investigation of the electronic band structure and optical properties of a two-dimensional anisotropic semimetal that is described by a tilted semi-Dirac type spectrum with a pair of Weyl nodes. We observe that a tilt along the quadratic direction can give rise to an energy imbalance between these nodes, contrary to the effect of tilt along the linear direction. We investigate the optical response of such system subjected to an external AC bias, aiming to probe the energy imbalance between the nodes. We show that the anisotropic interband optical conductivity gives a clear signature of imbalanced nodes by exciting electrons at two different chemical potentials at near zero frequency and the difference between these two chemical potentials is the direct measure of the energy imbalance. Subsequently, we also investigate the intraband DC conductivity by using the semi-classical Boltzmann transport theory which reveals that contrary to the tilted Dirac materials, tilt can convert semi-Dirac material from semimetallic phase to metallic phase. Furthermore, we periodically drive the system by external time-periodic perturbation to open up topological gap at those nodes. We also show that the presence of imbalanced Weyl nodes would prevent the semi-Dirac material from switching to Chern topological phase even after opening topological gaps at the nodes as the bulk remains gapless. Such state cannot be probed by the usual anomalous Hall response, as it will be overshadowed by the bulk contribution. Here, we show that those gaps at different chemical potentials can be probed by optical excitation.  Finally, we extend our study to nonlinear regime, where we particularly focus on second harmonic generation in an inversion symmetry broken tilted semi-Dirac system. A clear signature of energy imbalanced Weyl nodes can also be detected here.
\end{abstract}  
\maketitle
\section{\textbf {Introduction} }\label{sec:intro}
In recent times, two-dimensional ($2$D) electronic systems described by a semi-Dirac (SD) type band structure have received much attention owing to their peculiar electronic properties \cite{Dietl2008,Banerjee_pickett2009,kush2016,PhysRevX.14.041057,IpshitaKush}. The band structure of SD material is described by massless Dirac-like linear dispersion along one direction and massive parabolic dispersion along its orthogonal direction \cite{pardo_pickett2009,Montambaux2009,Banerjee_pickett2009,delpace}. The material was first proposed theoretically in 2008 in a deformed honeycomb lattice  \cite{Dietl2008} and has recently been realized experimentally \cite{PhysRevX.14.041057}. As noted in Ref.~[\onlinecite{delpace}], the lattice deformation induced weak asymmetry among the three nearest hoping parameters in a honeycomb lattice can cause an anisotropy in the band structure. In fact, an interesting situation can emerge when one among the three nearest-neighbor hoping parameters become double to another hopping parameter, i.e., $t_1=2t_2$,
for which two valleys merge at the $\Gamma$ point in the Brillouin zone and give rise to a SD type band with low-energy effective Hamiltonian $H=ak_x^2\sigma_x+v_Fk_y\sigma_y$, where $\sigma$'s are Pauli matrices in orbital space and $v_F$ is the Fermi velocity along the linear direction, $a$ is the inverse mass term, and ${\bf k}=\{k_x,k_y\}$ are usual $2$D momentum operators. Also, in the presence of longitudinal strain, this material mimics three-dimensional ($3$D) Weyl semimetal by splitting up the band-touching at $\Gamma$ point into two Weyl type nodes \cite{IpshitaKush}.

 The SD type band spectrum has received a significant attention from the theoretical front like investigating anisotropic diffusion transport \cite{Adroguer2016}, magnetoconductivity \cite{zhouchen2021,p_sinha}, direction sensitive optical conductivity \cite{Alestinmawrie,carbotte2019,OriekhovGusynin,Xiong2023,Ashutosh25}, photoinduced band structure modulation \cite{awdesh,SKF_Arijit,Chen,SKF_valkov}, unusual integer quantum Hall effects \cite{kush2016}, crossover from retro to specular Andreev refelction \cite{Li_2022}, anisotropic plasmon modes \cite{A_lurov}, magnetothermoelectric properties \cite{abedinpour}, interaction effect \cite{elsayed} etc. Additionally, with the presence of the two Weyl nodes, thermoelectric properties have been also carried out \cite{IpshitaKush}.  

 In two-band Dirac-like semimetals, sometimes lattice deformations induce a tilt to the Dirac cone, for example $8$-Pmmn borophene exhibits a pair of tilted Dirac cones at two valleys \cite{lozovik}. Such tilted Dirac cones have been also observed in $3$D Dirac-Weyl systems \cite{weylTaAS_2015,Deng_2016,weyltype2_2016}. Extensive theoretical and experimental works have been carried out revealing different exciting features associated with the tilted Dirac cones, such as tilt induced anisotropic plasmon excitation in $8$-Pmmn borophene \cite{amit2017,reza_2021}, particle-hole symmetry breaking and associated optical excitation \cite{alestin_borophene,mojarro} and valley polarized magnetotransport \cite{SKF_jayannavar,Islam_2018_jpcm}. On the other hand, tilted $3$D Weyl systems have also been studied extensively to investigate various aspects of tilted Dirac cones, from optical excitation \cite{xu_optic,Q_chen} to magnetotransport properties \cite{Vladmir_Zyuzin,Sharma_goswami,Wei_yi,Das_kamal}. Along the same line, tilt has been also considered in a $2$D SD system while investigating optical conductivity \cite{xu_yan2023}. However, tilt has been considered here along the linear direction, without any Weyl nodes.
 
  In this work, we aim to study the effect of tilt along the quadratic direction instead of linear direction, including the presence of two Weyl nodes. We first analyze the band structure and subsequently investigate the optical conductivity. We show that a unique case emerges here, an energy imbalance occurs between the two nodes which is quite similar to inversion symmetry broken $3$D Weyl semimetals \cite{zyuzin} and electron-electron interaction induced valley imbalance in graphene \cite{PhysRevB.100.121402}. We show a clear signature of optical conductivity at two different values of chemical potential, indicating the existence of energy-imbalanced Weyl nodes. We also discuss the DC conductivity so that the tilt can convert the SD material from semimetallic phase to metallic phase, contrary to the tilted Dirac material. Additionally, if such tilted SD material is periodically driven by external perturbation in the form of light, it will open topological gap at the two nodes. However, because of the energy imbalance between the two nodes, the chemical potential can never be adjusted inside the gap at both the nodes simultaneously, as a result, the anomalous Hall response would always be overshadowed by the bulk contribution. In this context, the optical response could be relatively more convenient than the DC electrical response to probe the system. We observed the gap signature as well as enhancement in the optical conductivity plots.
  
 We also look into the regime  beyond the linear response, the nonlinear optical response that has attracted an intense research interest recently in Dirac-Weyl system.  The nonlinear optical signatures have been known to exhibit several unique quantum phenomena as photovoltaic effect \cite{oka2009photovoltaic,dejuan2017quantized}, second harmonic generation (SHG) \cite{bykov2012second, wang2016giant, Xu2018SHGWeyl, sipe,khurgin,Bhalla2022,PhysRevB.95.155421} and third harmonic generation (THG) \cite{Cheng2014,Hong2013THGGraphene,Pilch2025THG} etc. In this work we particularly focus on the SHG  in which the electronic system with broken inversion symmetry absorbs light with frequency $\omega$ and emits with double the frequency $2\omega$.  
  We find that the presence of energy imbalanced Weyl nodes allows SHG to occur at two different chemical potentials around the nodes.
  
The manuscript is composed of six sections. After giving an introduction in the Sec.~(\ref{sec:intro}), we introduce the model Hamiltonian, corresponding band structure and density of states (DOS) of $2$D tilted SD system in Sec.~(\ref{sec:model}). We present the optical conductivity with discussion in Sec.~(\ref{sec:optical}). In Sec.~(\ref{sec:floquet}), we analyze the effect of external irradiation on the band structure and optical properties of tilted SD material. In Sec.~(\ref{sec:nonlinear}), we investigate the nonlinear optical response (SHG) of tilted SD with intrinsically broken inversion symmetry. Finally, we summarize our work in Sec.~(\ref{sec:conclusion}).
\section{\textbf {Model Hamiltonian and energy spectrum}}  \label{sec:model}
Let us consider that the SD material lies in the $x-y$ plane. The low-energy effective Hamiltonian is described by  $H={\bf d\cdot\sigma}$ where ${d_x}=
 (\alpha k_x^2-\delta_0)$ and ${d_y}=v_Fk_y$ \cite{Banerjee_pickett2009,Montambaux2009,kush2016}. Here, $\{\alpha,\delta_0\}$ are constant system parameters that measure the degree of lattice deformation and $v_F$ is the Fermi velocity along the y-direction. Also, $\{\sigma_x, \sigma_y\}$ are the Pauli matrices corresponding to pseudospin in orbital space and ${\bf{k}}=\{k_x, k_y\}$ are the $2$D momentum operators. The corresponding energy spectrum can be immediately obtained as $E_{{\bf k },\eta}=\eta \abs{d}$ with $\eta=\pm$ as band index. The energy spectrum is highly anisotropic, i.e., linear along $k_y$ and quadratic along $k_x$ direction. However, in this work we are interested in a tilted SD material, more specifically the tilt along the quadratic direction.
\begin{figure}[t]
    \centering
    \includegraphics[height=6cm,width=1\linewidth]{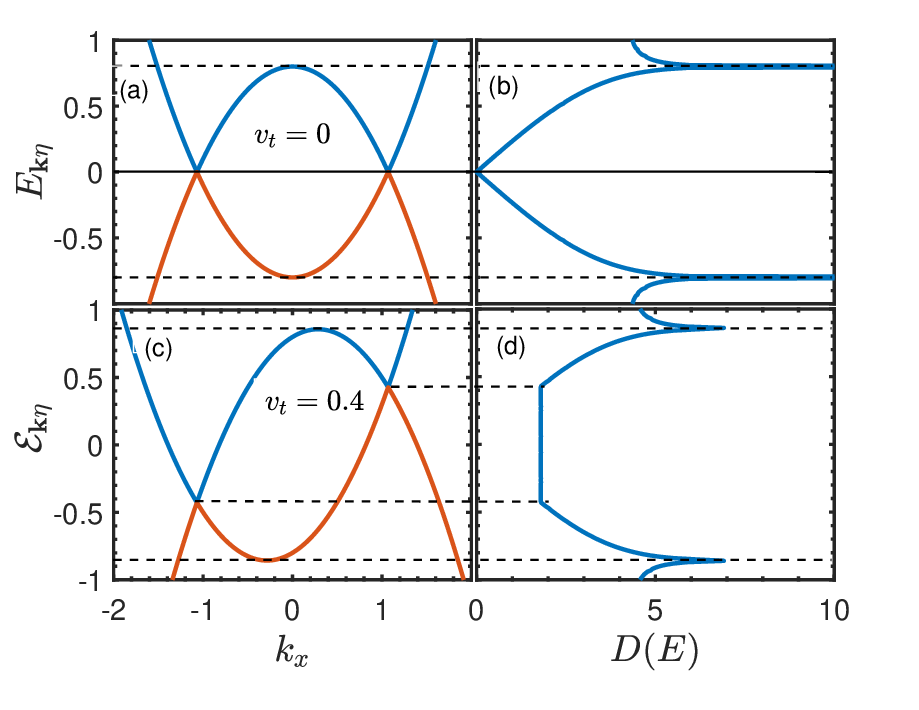}
    \caption{shows the band structure and corresponding DOS for non-tilted and tilted SD material. In  Fig.(a) and Fig.(b), we map the band structure and DOS for non-tilted SD material. While in Fig(c) and Fig.(d) tilted band structure is mapped to its DOS. The energy is taken in unit of $v_Fk_F$, $k_x$ in unit of $k_F$ and DOS in unit of $D_0$. Parameters used: $\delta_0=0.8$, $\alpha=0.7$, and $v_t=0.4$ in the unit of $v_Fk_F$ ,$v_F/k_F$ and $v_F$.}
    \label{fig:dos}
\end{figure}
The low-energy effective Hamiltonian for the tilted SD material is taken as $ \mathcal{H}={{d_0}\mathbb{1}}+H$, where $d_0=v_tk_x$ accounts the tilting along the $k_x$ direction with $v_t$ as the tilt parameter. The corresponding energy spectrum is given as $\mathcal{E}_{k,\eta}=d_0+\eta\abs{d}$.
As mentioned earlier, the presence of the tilting term induces an energy imbalance by $2v_t\sqrt{\delta_{0}/\alpha}$.

We now briefly discuss the effect of tilt on the density of states (DOS) which can be obtained as a function of energy as $D(E)=\displaystyle\sum_{\bf{k},\eta}\delta(E-\mathcal{E}_{k,\eta})$. By using delta function decomposition, we can write the DOS at energy $E$ as 
 \begin{equation}\label{dds}
 D(E)=D_0\int d k_x\frac{\abs{E-d_0}}{\sqrt{(E-d_0)^2-d_x^2}}\Theta{[(E-d_0)^2-d_x^2]}
 \end{equation} 
 where $D_0=k_F/2\pi^2v_F$, $\Theta(..)$ is Heaviside step function, and the $x$-component of momentum is normalized as $k_x\rightarrow k_x/k_F$ with $k_F$ being the typical Fermi momentum wave vector for a standard $2$D electronic system. The band structure of tilted and non-tilted SD material along the quadratic direction (at $k_y=0$) and the numerically computed corresponding DOS by using Eq.~(\ref{dds}) are presented in Fig.~\ref{fig:dos}. In the case of non-tilted SD material, the DOS increases almost linearly with energy and attains a peak exactly at the point where $\partial E/\partial k_x=0 $ (see Fig.~\ref{fig:dos} (b)). While in tilted case, the DOS exhibits a non-zero constant value in between the two Weyl nodes in energy space, and then attains peaks at the band edges before decreasing again. It is because in the tilted case, when the chemical potential is placed at zero, two nodes of the SD system behave differently and contribute to the non-zero DOS. The constant DOS between the two nodes can be attributed to the fact that when the chemical potential is varied from one node to another in energy space, the individual contributions from these two nodes cancel each other, resulting in a constant DOS.  Further, the presence of two kinks at $\pm v_t\sqrt{\delta_0/\alpha}$ corresponds to the two imbalanced nodes as seen in Fig.~\ref{fig:dos}(d).
\section{\bf  Optical Conductivity}\label{sec:optical}
Let us consider that the tilted SD material is subjected to an external in-plane AC electric field $\mathbf {E}(\mathbf{r},t)=\mathbf{E_0}(\mathbf{r})e^{i\omega t}$ and the frequency is tunable externally. Here, $E_0$ is the amplitude of the bias. The response to this field can be seen via optical conductivity, which can be computed by using linear response theory. However, as the electronic band structure is anisotropic in nature, the optical response is also expected to be anisotropic, i.e., the response is sensitive to the direction of the applied field. Our particular focus will be on the interband electron excitation. We will also discuss the intraband transport. The electron from the valence band absorbs an energy $\omega$ from the applied AC field and makes a transition to the conduction band through chemical potential. Similarly, it can emit the same energy and relax to the valence band. This phenomenon can be probed via the real part of interband optical conductivity, making it a crucial tool to probe the band structure of material.

 In order to compute the optical conductivity, we shall use the well-known Kubo formula based on linear response theory \cite{Mahan2000}. Following this, we can write the different components of the optical conductivity as
\begin{multline}   \label{eq:kuboformula}
    \sigma_{\alpha\beta}=i\frac{e^2}{\omega}\int\frac{d^2k}{(2\pi)^2} { T\sum_cTr\langle \hat v_{\alpha} \widehat{\mathcal{G}}(\mathbf{k},\omega_c)}\\
    \times \hat v_{\beta} \widehat{\mathcal{G}}(\mathbf{k},\omega_c+\omega_d)\rangle_{i\omega_d\rightarrow\omega+i\zeta_o}\
\end{multline}  
with $\{\alpha,\beta\}$ represent the components of spatial coordinate $\{x, y\}$. The temperature is denoted by $T$ and $\omega_{c,d}$ denotes the Matsubara frequency. The different components of velocity operators are $\hat v_{x}=\partial{\mathcal{H}}/\partial{k_x}=v_{t}\mathbb{1}+2\alpha k_x\sigma_x$ and $\hat v_y=\partial{\mathcal{H}}/\partial{k_y}=v_F\sigma_y$. 
The Matsubara Green function for the $2$D tilted SD material is obtained as
\begin{equation}  \label{eq:greensfn}
  \widehat{\mathcal{G}}({\bf k},\omega)=\frac{1}{2}\sum_\eta \frac{\mathbb{1}+\eta{\hat{d }\cdot\sigma} }{i\omega+\mu-{\mathcal{E}}_{\bf k,\eta}}
\end{equation}
where $\hat d={\bf d}/d$. Now, substituting Eq.~(\ref{eq:greensfn}) into the Eq.~(\ref{eq:kuboformula}) and performing the following Matsubara summation
\begin{widetext}
 \begin{align} \label{eq:mastur}
T \sum_c \frac{1}{i\omega_c + \mu - \mathcal{E}_{\mathbf{k}, \eta}} 
          ~ \frac{1}{i(\omega_c + \omega_d) + \mu - \mathcal{E}_{\mathbf{k}, \eta'}} 
= 
\begin{cases}
\displaystyle \frac{f(\mathcal{E}_{\bf k,\eta}) - f(\mathcal{E}_{\bf k, \eta'})}
               {i \omega_d- \mathcal{E}_{\bf k,\eta'} + \mathcal{E}_{\bf k,\eta}}, & \text{for } \eta \neq \eta' \\
0, & \text{otherwise}
\end{cases}
\end{align}
 we arrive at
\begin{align}
     \sigma_{xx}=i\frac{e^2}{\omega} 2 \sum\limits_{\eta\eta'} \int\frac{d^2k}{(2\pi)^2} \Bigg[
        \alpha(d_x+\delta_0) \left( 1 + \eta \eta' \frac{d_x^2 - d_y^2 }{d^2} \right)
        + v_{t}^2 \left( 1 + \eta \eta' \right)
        + \frac{4\alpha (\eta + \eta')  d_0 d_x}{\abs{d}} 
         \Bigg] 
         \frac{f\mathcal{(E}_{\bf k,\eta}) - f(\mathcal{E}_{\bf k, \eta'})}{\omega+i \zeta_o - \mathcal{E}_{\bf k, \eta'} + \mathcal{E}_{\bf k,\eta}}.
\end{align}
Here, $f(\mathcal{E_{\bf k,\eta}})=\big[\exp[(\mathcal{E_{\bf k,\eta}}-\mu)/k_BT]+1\big]^{-1}$ is the Fermi-Dirac distribution function with $\mu$ being the chemical potential.

\subsection{\bf Interband optical conductivity} \label{sec:A}
The real part of the interband optical conductivity corresponds to the optical absorption and arises from the interband transitions through the chemical potential, can be written for $\eta\ne\eta'$ as
\begin{align}
\operatorname{Re}(\sigma_{xx})=\frac{e^2}{\omega} 2\pi \sum\limits_{\eta\eta'}\int\frac{d^2k}{(2\pi)^2} \Bigg[
        \alpha(d_x+\delta_0) \left( 1 + \eta \eta' \frac{d_x^2 - d_y^2 }{d^2} \right)
        + v_{t}^2 \left( 1 + \eta \eta' \right)+\frac{4\alpha (\eta + \eta') d_0 d_x}{\abs{d}}  \Bigg]\nonumber\\ 
        \times\left[{f(\mathcal{E}_{\bf k, \eta}) - f(\mathcal{E}_{\bf k, \eta'})}\right]\delta(\omega - \mathcal{E}_{\bf k, \eta'} + \mathcal{E}_{\bf k, \eta}).
 \end{align}
 Note that the intraband excitation is suppressed here as it requires momentum transfer that is negligibly small in a clean system, which can also be seen from Eq.~(\ref{eq:mastur}). Hence, we proceed with the interband transition only by setting $\eta=-$ and $\eta'=+$ as 
\begin{align}
\operatorname{Re}(\sigma_{xx})=
\frac{e^2}{\omega} 4\pi \int\frac{d^2k}{(2\pi)^2}
\alpha(d_x+\delta_0)\Bigg( \frac{d_y}{d} \Bigg)^2  \left[f(\mathcal{E}_{\bf k,-}) - f(\mathcal{E}_{\bf k,+})\right]\delta(\omega - 2\abs{d}),
\end{align}
which can be further simplified by using the delta function decomposition as
\begin{align} \label{eqn:sigma_xx}
\operatorname{Re}(\sigma_{xx})=\frac{2e^2}{\pi} \frac{1}{v_F\omega^2} \int \mathrm{d}{k_x}~ 
        \alpha(d_x+\delta_0) \sqrt{\bigg(\frac{\omega}{2}\bigg)^2 - d_x^2}
        ~\Gamma(k_x,\omega,T,\mu)~\Theta\bigg[\bigg(\frac{\omega}{2}\bigg)^2-d_x^2 \bigg],
\end{align}
where
\begin {align}
\Gamma(k_x,\omega,T,\mu)=\frac{\sinh(\beta\omega/2)}{\cosh(\beta\omega/2)+\cosh(\beta(v_tk_x-\mu))}.
\end{align}
 In the similar way, the real part of $yy$-component of interband optical conductivity is obtained as
\begin{align} \label{eqn:sigma_yy}
\operatorname{Re}(\sigma_{yy})=\frac{e^2}{2\pi} \frac{v_F}{\omega^2} \int\mathrm{d}{k_x} \frac{d_x^2 
         }{\sqrt{(\omega/2)^2-d_x^2}} ~\Gamma(k_x,\omega,T,\mu)~\Theta\bigg[\bigg(\frac{\omega}{2}\bigg)^2-d_x^2\bigg].
\end{align}
\begin{figure*}[t]
 \centering
 \subfigure[]
 {\label{fig:sig_xx}\includegraphics[width=0.48\linewidth]{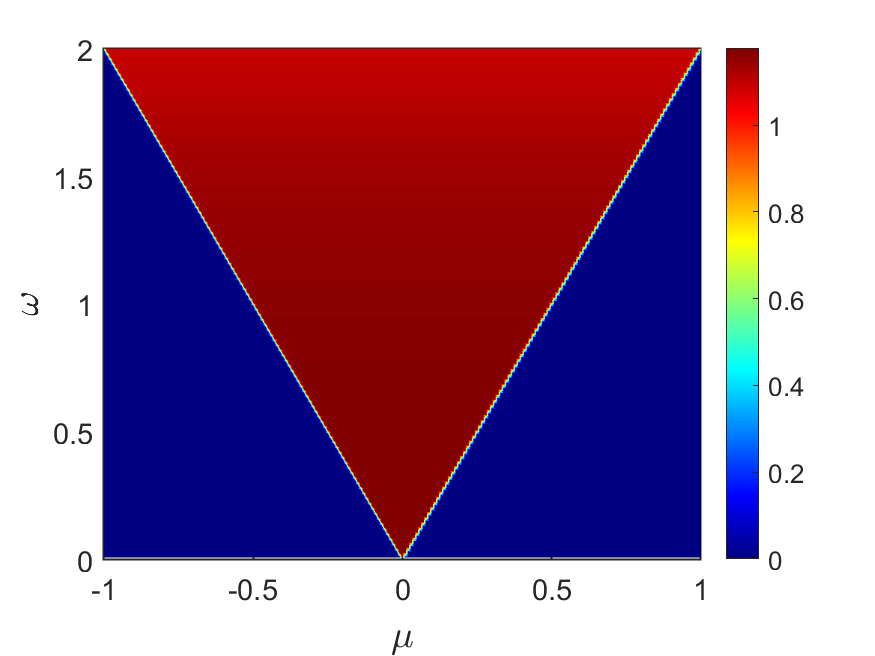}}
 \subfigure[]
 {\label{fig:sig_yy}\includegraphics[ width=0.48\linewidth]{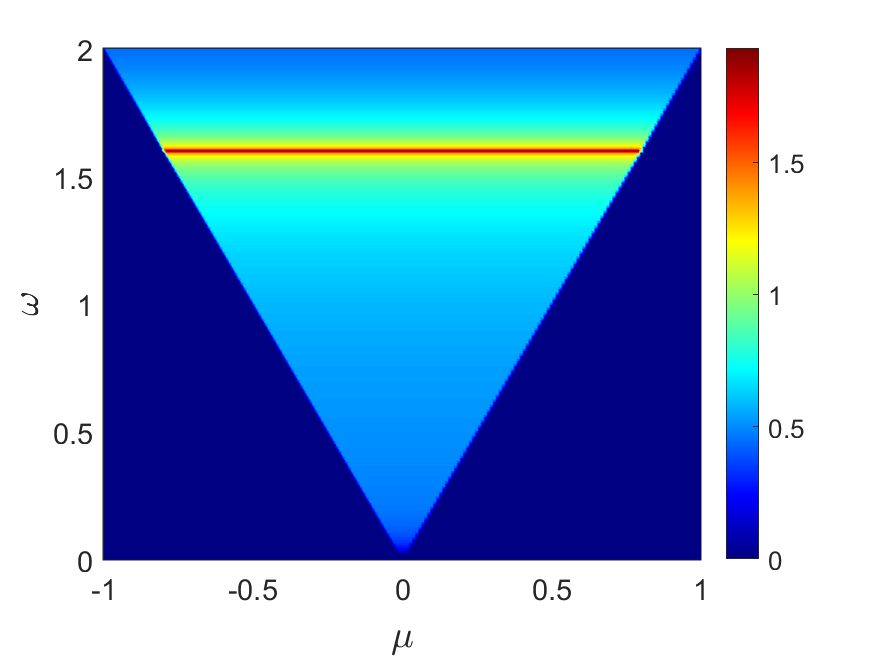}}
 
 \subfigure[]
{\label{fig:tsigma_xx}\includegraphics[width=0.48\linewidth]{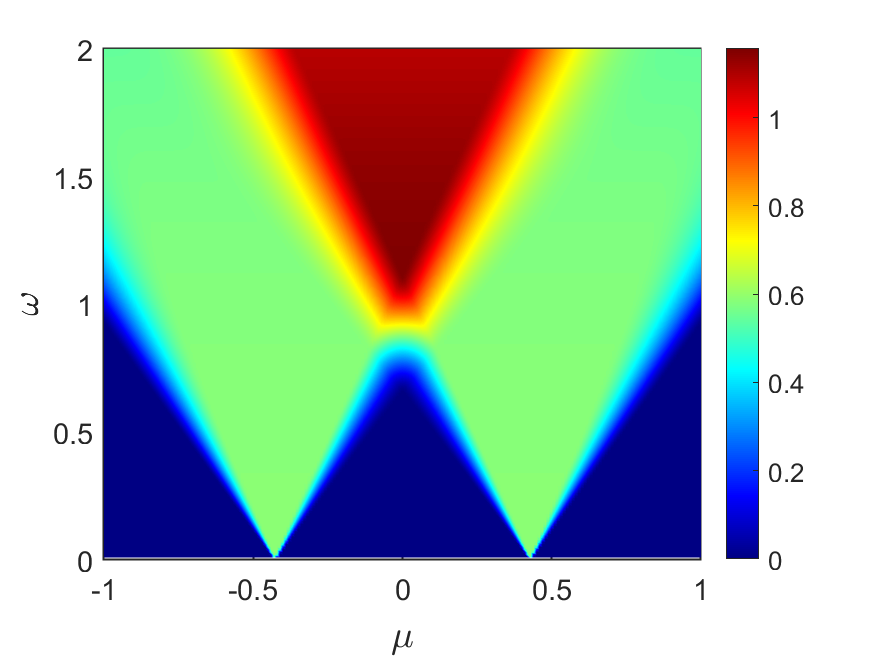}}
 \subfigure[]
 {\label{fig:tsigma_yy}\includegraphics[width=0.48\linewidth]{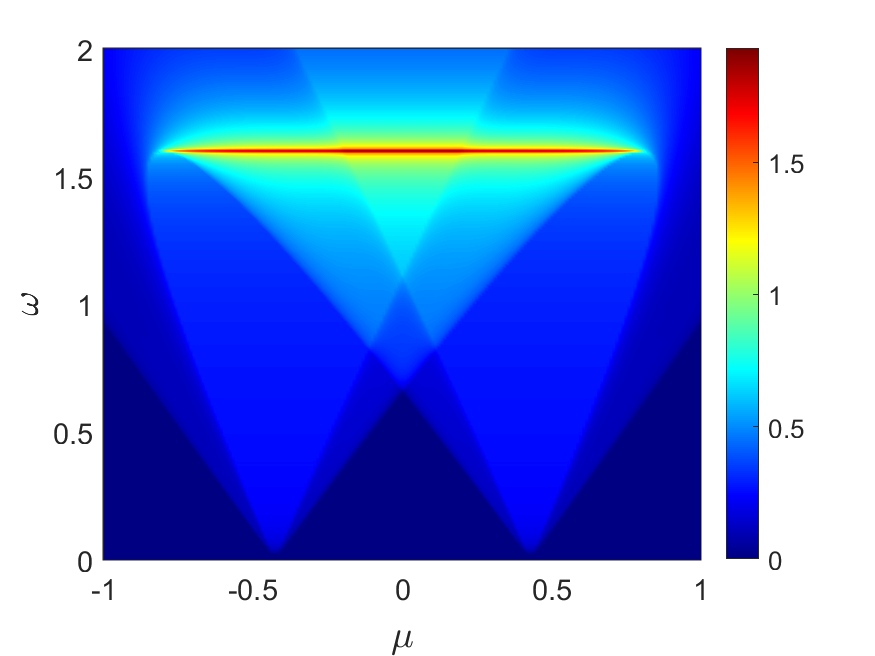}}
 \caption{Contour plot of  $\operatorname{Re}[\sigma_{xx}(\mu,\omega)]$ and $\operatorname{Re}[\sigma_{yy}(\mu,\omega)]$ for non-tilted and tilted SD material. For non-tilted SD material, the $\operatorname{Re}[\sigma_{xx}(\mu,\omega)]$ and $\operatorname{Re}[\sigma_{yy}(\mu,\omega)]$ are shown in Fig.~\ref{fig:sig_xx} and \ref{fig:sig_yy}. while as in the case of tilted SD material, the $\operatorname{Re}[\sigma_{xx}(\mu,\omega)]$  and $\operatorname{Re}[\sigma_{yy}(\mu,\omega)]$ are shown in \ref{fig:tsigma_xx} and \ref{fig:tsigma_yy} respectively. Each subplot has a color bar that represents the magnitude of optical conductivity in the unit of $e^2/2\pi$.}
 \label{fig:heat_map_conductivity} 
 \end{figure*}
 \end{widetext}
 
Note that the anisotropy enters the optical conductivity through different components of the velocity matrix operators $\hat{v}_x$ and $\hat{v}_y$ which stem from the anisotropic band structure. From the above expression, it can be seen that only those electrons with $k_x\in[-\sqrt{(\omega/2+\delta_0)/\alpha},\sqrt{(\omega/2+\delta_0)/\alpha}]$ are excited and its range increases with the increase of the separation between the two nodes $\delta_0$.

First, we briefly discuss the non-tilted case. The Fig.~\ref{fig:sig_xx}
depicts the behavior of Re[$\sigma_{xx}]$ without any tilt. In this case, the two Weyl nodes are degenerate and sit at the zero energy level. Hence, the optical excitation is turned on around both nodes simultaneously at near zero frequency ($\omega\sim0$) for undoped situation ($\mu=0$). It is significantly very high at near zero frequency, and it slowly decreases with the further increase of the frequency. Note that the behavior of optical conductivity is directly related to the DOS which enters through momentum integration, as well as the direction dependent velocity matrix. For doped case $\mu\ne 0$, the optical transitions occur for $\omega\ge2\mu$ when an electron from the valence band below the chemical potential receives energy $\omega$ from the AC bias and is excited to the conduction band above the chemical potential. We plot Re[$\sigma_{yy}$] in the Fig.~\ref{fig:sig_yy} which shows that conductivity increases slowly with frequency and similar to the previous case, here also electron excitation starts even at near zero frequency for undoped case. However, the optical conductivity exhibits a sharp peak around $\omega\simeq 2\delta_0$ (separation between the band edges at $\Gamma$ point) as denoted by a red horizontal line in the Fig.~\ref{fig:sig_yy} which was observed in Ref.~\cite{Alestinmawrie} and termed as giant optical conductivity. The origin of such giant peak can be attributed to the sharp increase of the DOS around the band edge at $\Gamma$ point, as shown in Fig.~\ref{fig:dos}.

Now we turn on the tilt parameter and plot the Re$[\sigma_{xx}]$ in the Fig.~\ref{fig:tsigma_xx}. Here, it is observed that the optical excitation starts at near zero frequency for two different values of chemical potential indicating the existence of energy-imbalanced Weyl nodes. The tilt parameter can be immediately estimated by setting $\mu_c=v_t\sqrt{\delta_0/\alpha}$ provided that $\delta_0$ is known, here $\mu_c$ is the chemical potential for which optical excitation starts at near zero frequency. Note that the strength of optical conductivity across each node is half of the non-tilted case, as expected because of the removal of node degeneracy. However, we note that when $\mu$ lies between the two nodes and the frequency of applied bias exceeds the limit $\omega>2\delta_0$ the optical conductivity sharply increases as shown by the triangular-shaped red region in Fig.~\ref{fig:tsigma_xx} that is because of the higher DOS around the band edge.

On the other hand, the $y$-component of optical conductivity $Re[\sigma_{yy}]$ in Fig.~\ref{fig:tsigma_yy} for tilted case still exhibits the giant optical conductivity at $\omega=2\delta_0$ except the range of chemical potential is now limited contrary to the non-tilted case. Additionally, we also observe that optical excitation occurs at each node individually as expected. The power law for the frequency dependency of optical transition may not be exactly solvable, but under two limiting cases $\omega\gg \delta_0$ and $\omega\ll\delta_0$ an analytical approximate form is briefly discussed in the Appendix \ref{power_law}.

\subsection{\bf Intraband Drude conductivity at $\omega\rightarrow 0$ } 
In this section, we discuss the usual DC conductivity, which describes the intraband transport in the zero-frequency limit. For the intraband DC transport we can obtain from Eq.~(\ref{eq:kuboformula}) at $\omega\rightarrow 0$
\begin{align}  \label{eqn:drude_conductivity}
\sigma_{\alpha\alpha}^{D}=e^2\int\frac{d^2k}{(2\pi)^2}|\langle\hat{v}_{\alpha}\rangle|^2\left[-\frac{\partial f(E)}{\partial E}\right]
\end{align}
where, $\hat{v}_{\hlm{\alpha}}$ denotes the $\hlm{\alpha}$-th component of velocity operator and $\langle\hat{v}_{\hlm{\alpha}}\rangle$ denotes the expectation value of $\hat{v}_{\hlm{\alpha}}$ with respect to the eigen states of Hamiltonian. The different velocity components for the tilted SD material are obtained as $\la v_{x}\ra=v_{t}+\eta~2 \alpha k_x d_x/\abs{d}$ and $\la v_y \ra=\eta 2v_Fk_y/\abs{d}$. As the system does not possess any topological gap, no anomalous Hall response is expected, i.e., $\sigma_{xy}=0$. Hence, we proceed with only longitudinal conductivity.
At the limit $T\rightarrow 0$, by substituting $-\partial f/\partial E= \delta \left[E-\mu\right]$ into Eq.~\eqref{eqn:drude_conductivity} we obtain the $x$-component of the Drude conductivity as
\begin{multline}   
\sigma_{xx}^D = \frac{e^2}{2\pi^2v_F}\int \mathrm{d}k_x~
\frac{\left(v_t \abs{\mu - d_0} + \eta~2\sqrt{\alpha(d_x+\delta)}d_x\right)^2}
{\abs{\mu - d_0} \sqrt{(\mu - d_0)^2 - d_x^2}}\\\times\Theta\left[(\mu-d_0)^2-d_x^2\right]
\end{multline}
and $y$-component as
\begin{align}
\sigma_{yy}^D=\frac{e^2v_F}{2\pi^2}\int {dk_x}\frac{\sqrt{(\mu-d_0)^2-d_x^2}}{\abs{\mu- d_0}}~\Theta\big[(\mu-d_0)^2-d_x^2\big].
\end{align}

\begin{figure}[t]
\centering
\begin{minipage}[t]{0.5\textwidth}
\hspace{-0.4cm}
\subfigure[]
{\label{fig:drude_xx}\includegraphics[height=4.5cm,width=0.53\linewidth]{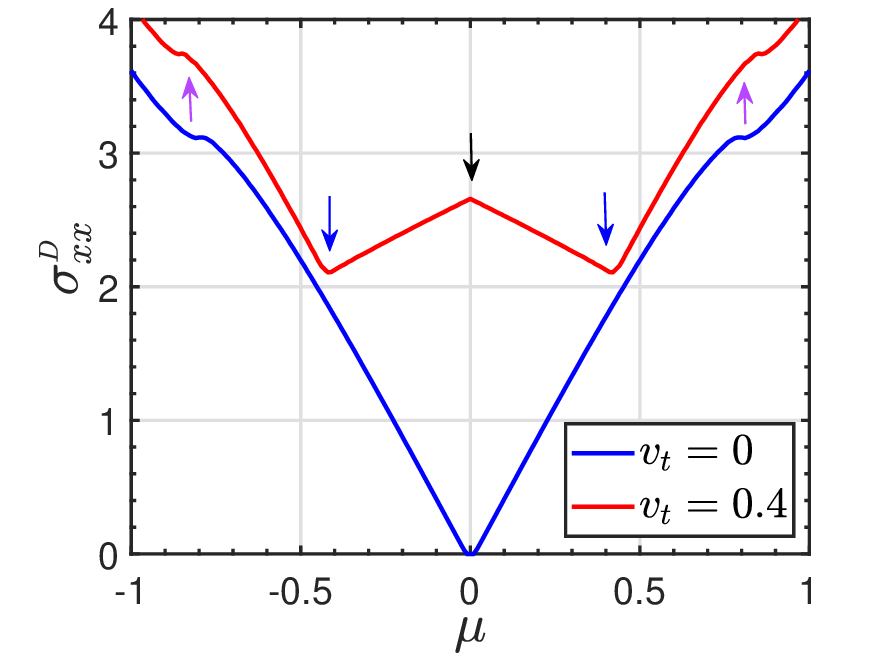}}
\hspace{-0.55cm}
\subfigure[]
{\label{fig:drude_yy}\includegraphics[height=4.5cm,width=0.53\linewidth]{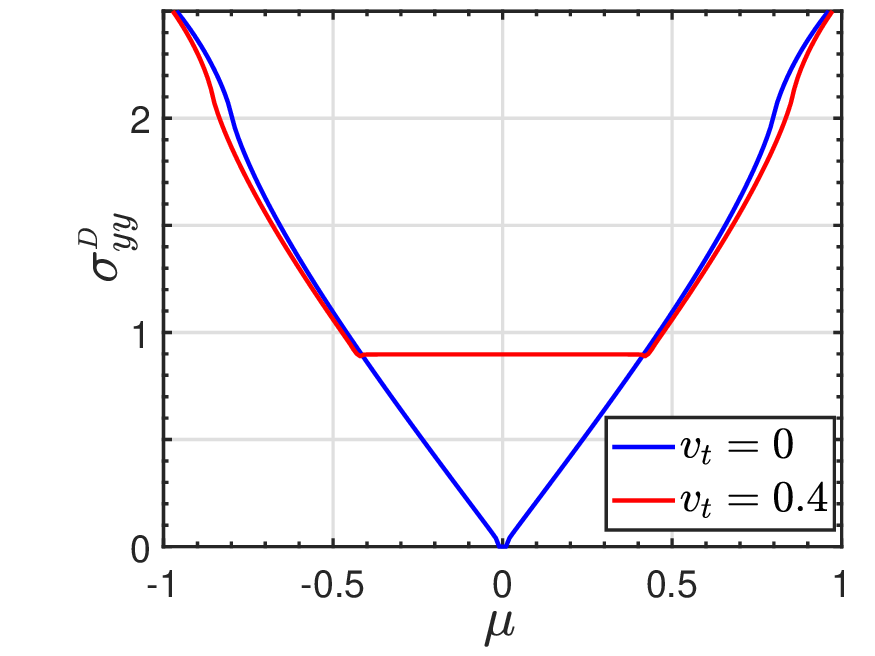}}
\end{minipage}
   \caption{The Drude conductivity (in units of $e^2/2\pi^2$) versus chemical potential (in units of $v_F k_F$) for x-component [Fig.~\ref{fig:drude_xx}] and y-component [Fig.~\ref{fig:drude_yy}] . The blue and red curves correspond to tilted and non-tilted SD material respectively.}
   \label{fig:drude}
\end{figure}
The above two equations are computed numerically and plotted in the Fig.~\ref{fig:drude_xx} and Fig.~\ref{fig:drude_yy}. The $x$-component of the Drude conductivity for non-tilted case vanishes for $\mu=0$ and keeps increasing almost linearly with doping. It is expected as the system has zero DOS at undoped case, as can be seen in Fig.~\ref{fig:dos}(b). However, there is very high DOS at the band edges, the conductivity remains almost unaffected except a small shift towards higher $\mu$  (as indicated by the vertical purple arrows). It is because without the tilt, $\langle \mathbf v\rangle$ vanishes at $\Gamma$ point, i.e., those highly dense electrons are strongly localized and do not participate in the DC transport. In presence of slight tilt $v_t\ne0$, the Drude conductivity becomes non-zero even at $\mu=0$. This is because the DOS and group velocity are now non-zero in presence of the tilt, as the group velocity is given by $v_x=v_t\pm 2\alpha k_x^{F}d_x/\abs{d}$ where $k_x^{F}$ is the solution of $|E_{k}|=0$. The Drude conductivity acquires a peak at $\mu=0$ as indicated by black arrow, and with the increase of doping, it attains a minima while passing through one node. This is because, the group velocity at the node point, i.e., $v_x=v_t$ as $d_x=0$ attains minima. The separation between two minima is the energy imbalance between two nodes. Here, we shall mention that in a  Dirac material with tilted Dirac cones- like $8$-Pmmn borophene where two Dirac cones are tilted in opposite direction, the tilt does not induce any conductivity to the undoped system as the DOS at $\mu=0$ remains zero \cite{Alestinmawrie}. Hence we can conclude that the tilt can convert an undoped SD system, with two Weyl nodes, from a semimetallic state to trivial metallic phase. 

Now, we look into the $y$-component of the Drude conductivity, plotted in Fig.~\ref{fig:drude_yy}. It shows that $\sigma_{yy}^D$ starts increasing linearly from zero, as expected because of the linear dispersion along the $y$-direction. However, with the tilt the uniform DOS is directly reflected here, the two nodes in energy space are emerging in Drude conductivity here. It is because the DOS plays the dominant role in $\sigma_{yy}^D$ whereas $|\langle v_x\rangle|^2$ overshadows the behavior of DOS in $\sigma_{xx}^D$.

\section{ \bf Floquet engineering SD system} \label{sec:floquet}
In this section, we aim to analyze the band structure of the periodically driven tilted SD material with two Weyl nodes, and subsequently explore its optical response. The band structure of a periodically driven SD material has been previously investigated without any tilt in the context of band topology and interfacial modes \cite{SKF_valkov,kush2016}.
Here, we briefly present the irradiated band structure of the SD system with tilt along the quadratic direction.

\subsection{Effective Hamiltonian and irradiated band structure}
We consider the periodic drive in the form of external light or irradiation propagating along z-direction represented by the vector field $\mathbf{A}(t)=A_0[\cos(\Omega t),\cos(\Omega t+\phi)]\hat{k}$, where $A_0$ is the amplitude, $\Omega$ and $\phi$ are frequency and phase respectively. The effect of this light field can be included to the unperturbed low-energy effective Hamiltonian via canonical momentum, i.e., $\mathbf{k}\rightarrow \mathbf{k}+e\mathbf{A}(t)$, $e$  being an electronic charge here.

The periodically driven Hamiltonian can be solved by using Floquet theory \cite{Eckardt2017}. We shall restrict our discussion to the low-energy regime for which it is sufficient to proceed with the effective Hamiltonian obtained by using the high frequency limit. In this limit, we use Floquet-Magnus expansion in the powers of  $1/\Omega$ as $H( {\bf k},t)\simeq H+H_F^{(0)}+H_F^{(1)}+\ldots$, where $H_F^{(0)}$ is time independent correction term, and  $H_{F}^{(1)}$ is first order correction term which can be solved as
$H_F^{(1)}=[H_{-},H_{+}]/{\Omega}$ 
where, $H_\pm=\int_0^{T}\mathcal{V}(t)e^{\pm i\Omega t}dt/T$.

 \begin{figure}[t]
     \centering
     \includegraphics[width=0.99\linewidth]{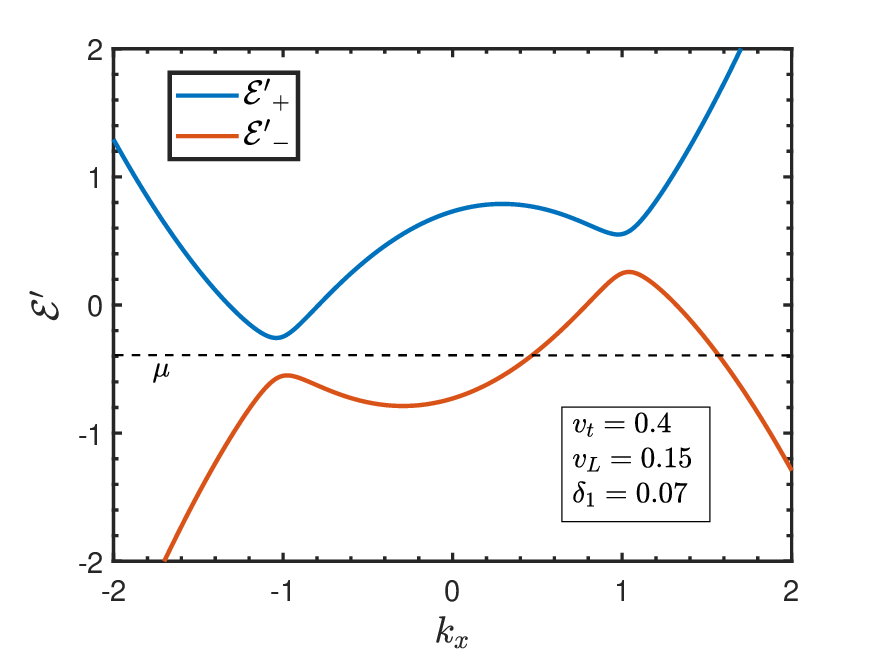}
     \caption{Band structure of irradiated tilted SD material along $k_x$ (at $k_y=0$) }
     \label{fig:light_band}
 \end{figure}   
\begin{figure*}[t]
\centering
\subfigure[]
{\label{fig:tgsigmaxx}\includegraphics[width=0.329\linewidth]{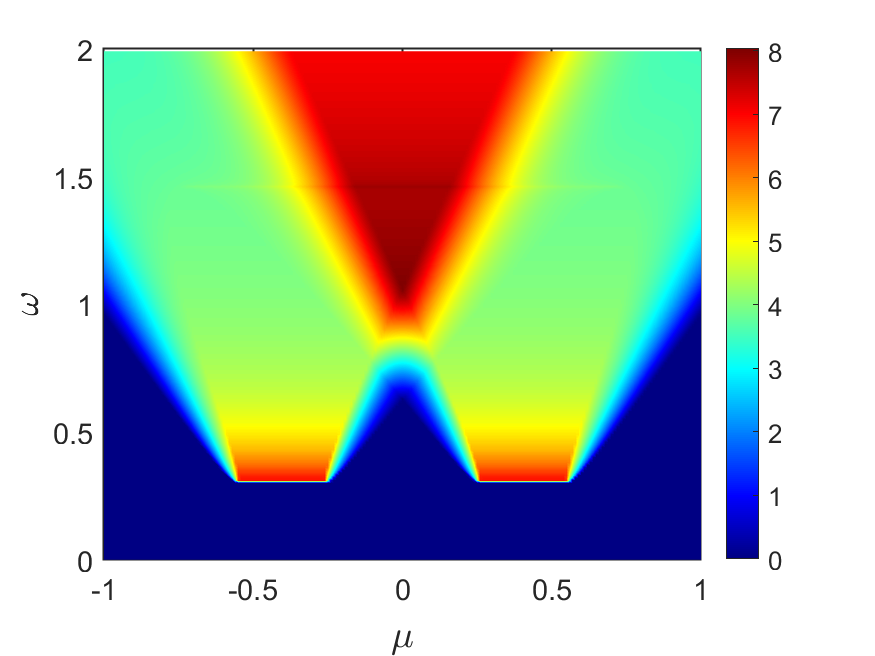}}
\centering
\subfigure[]
{\label{fig:tgsigmayy}\includegraphics[width=0.329\linewidth]{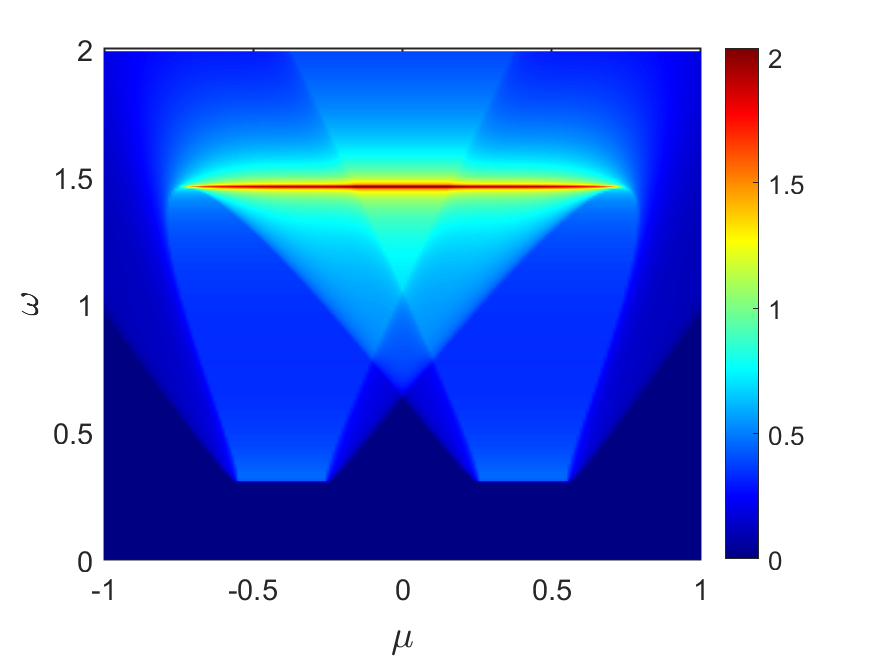}}
\centering
\subfigure[]
{\label{fig:tgsigmaxy}\includegraphics[width=0.329\linewidth]{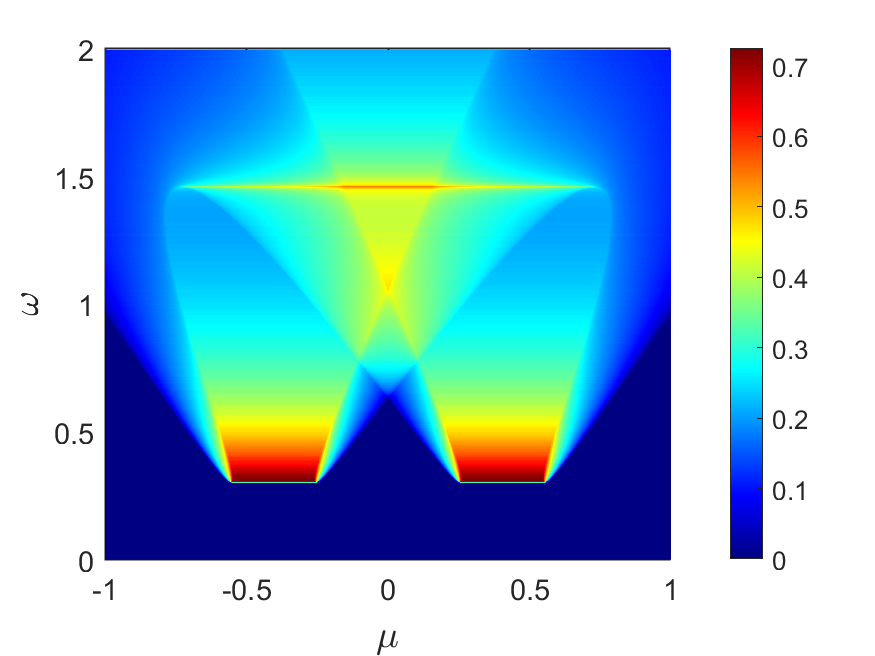}}

\caption{Contour plots showing the real part of the interband optical conductivity for a circularly polarized irradiated tilted SD material, as a function of photon frequency $\omega$ and chemical potential $\mu$. The color scale represents the magnitude of (a) $\mathrm{Re}[\sigma_{xx}(\omega, \mu)]$, (b) $\mathrm{Re}[\sigma_{yy}(\omega, \mu)]$ and (c) $\mathrm{Im}[\sigma_{xy}(\omega, \mu)]$, in units of $e^2/2\pi$.}
\label{fig:irr_optical cond.}
\end{figure*}
 
For our system, $H_F^{(0)}=\alpha e^2A_0^2\sigma_x/2$ and, $ \mathcal{V}(t)={\bf \mathcal{W}(t)\cdot\sigma} $ where $\mathcal{W}_0=v_{t} eA_0\cos(\Omega t)$, $\mathcal{W}_x(t)=2\alpha eA_0 k_x\cos(\Omega t)-\alpha e^2 A_0^2\cos(2\Omega t)/2$ and $\mathcal{W}_y(t)=ev_F A_0\cos(\Omega t+\phi)$. 
The first-order correction term is obtained as $H_{F}^{(1)}=v_{L} k_x\sigma_z$ with $v_{L}=(\alpha v_F e^2 A_0^2/\Omega)\sin\phi$, this correction term changes sign at the two nodes. So, the effective Hamiltonian in presence of irradiation can be written as $\mathcal{H}_{eff}=d_0\mathbb{1}+\bf d'\cdot\sigma$ where, $d'_x=(\alpha k_x^2-\delta')$, $d_y'=v_F k_y$, $d_z'=v_L k_x$ and $\{\sigma_x,\sigma_y,\sigma_z\}$ are Pauli matrices in orbital space. The corresponding Floquet energy spectrum is obtained as $\mathcal{E'}_{k,\eta}=d_0+\eta\abs{d'}$ which is plotted in the Fig.~\ref{fig:light_band}. The term $d_z'$ is the momentum dependent gap term and $\delta'=\delta_0-\delta_1$ with $ \delta_1=\alpha e^2 A_0^2/2$.  This momentum dependent gap term ($d'_z$) breaks the time-reversal symmetry (TRS) and opens up the topological gap $2|v_L\sqrt{\delta'/\alpha}|$ at the two node points $k_x^{\pm}=\pm\sqrt{\delta'/\alpha}$, provided $0<\phi< \pi$. Generally such photoinduced topological gap in two band semimetal-like graphene \cite{McIver2020} was reported to be around $10-100$ meV when driven by circularly polarized light with field strength $10^7\sim 10^8$ V/m. 

Now, we turn to Chern number given as $\mathcal{C}=\int_k\mathcal{B}(k)$, where $\mathcal{B}(k)$ is the Berry curvature. We proceed by the usual two-band Berry curvature formula and obtain $\mathcal{B}(k)=\pm v_Lv_F(d_x+2\delta)/2(d_x'^2+d_y'^2+d_z'^2)^{3/2}$, here we observe $\mathcal{B}(k)\ne -\mathcal{B}(-k)$ which leads to $\mathcal{C}=\pm1$. This has been already predicted in the Ref. \cite{kush2016}.
Note that the situation is very unique, although light opens up topological gap at both the nodes, the system can never attain a bulk insulating phase even though non-zero Chern number indicates the presence of topologically protected edge modes. It can also be clearly seen that if the chemical potential is placed inside the topological gap at one node, the other node remains doped ( see Fig.~\ref{fig:light_band} ), hence the system will always be at bulk conducting phase. In such case, the anomalous Hall response will be always overshadowed by the bulk contribution. Hence, an optical probe of such phase is more convenient than DC transport.

Note that $\delta_1$ term that arises from the light is not sensitive to the light polarization whereas the topological gap parameter associated with $\sigma_z$ in the effective Hamiltonian appears only when $\phi\neq (0,\pi$). The linearly polarized light can induce $\delta_1$ term even without opening a gap. In the presence of light, the positions of the nodes are at $k_x^{\pm}=\pm\sqrt{(\delta_0-\delta_1)/\alpha}$, indicating that inter-node momentum separation can be reduced by applying linearly polarized light. In fact, one can easily estimate the value of $\delta_0$ by just adjusting $\delta_1$, for example two nodes merge at $k_x=0$ for $\delta_1=\delta_0$.
 
 \subsection{Interband optical conductivity in irradiated SD system} \label{subsec:IVB}  The effects of irradiation on the optical conductivity are briefly presented here. We follow the similar approach as in Sec.~\ref{sec:A}. Here, we shall mention that in general one has to set-up the Kubo formalism based on Floquet Green function which requires summation over all Floquet side bands \cite{ali_Sadeghi}. However, as we are mainly interested near the $\mu=0$, we can safely ignore the effects of higher side bands and proceed with the effective Hamiltonian and corresponding Green function. Using effective Hamiltonian, we arrive at the expression for the real part of longitidinal optical conductivity, as follows
 \begin{widetext}
 \begin{eqnarray}  \label{eq:fsigmaxx}
 \operatorname{Re}({\sigma_{xx}}) &=& \frac{e^2}{2\pi} \frac{1}{v_F \omega^2} \int \mathrm{d}k_x\;
\frac{4\alpha(d_x'+\delta')\left[(\omega/2)^2 - d_x'^2 \right] + v_L\left[(\omega/2)^2 - d_z'^2\right] - 4\alpha d_x'd_z'^2}
{\sqrt{(\omega/2)^2 - d_x'^2 - d_z'^2}}
\Gamma(k_x,\omega,T,\mu)\nonumber\\&\times&\Theta\left[(\omega/2)^2 - d_x'^2 - d_z'^2\right],
  \end{eqnarray}
and the $y$-component as 
\begin{align}    \label{eq:fsigmayy}
   \operatorname{Re}(\sigma_{yy})=\frac{e^2}{2\pi} \frac{v_F}{\omega^2} \int\mathrm{d}{k_x} \frac{d_x'^2 +d_z'^2
         }{\sqrt{(\omega/2)^2-d_x'^2-d_z'^2}} ~\Gamma(k_x,\omega,T,\mu)~\Theta\bigg[\bigg(\frac{\omega}{2}\bigg)^2-d_x'^2-d_z'^2\bigg].
\end{align}   
Now we turn to the off-diagonal components of the optical conductivity tensor. Because of the Floquet engineered topological gap, a nonvanishing dynamical Hall response $\sigma_{xy}$ is also expected. As mentioned previously,  the tilt does not allow the bulk to be fully gapped even though topological gaps open up at the nodes. Hence, an usual DC quantum anomalous Hall conductivity will be always overshadowed by the bulk contribution rather an AC Hall conductivity could be more convenient to probe such topological gap. In contrast to the longitudinal components, where the real part governs the absorptive response, the imaginary part of $\sigma_{xy}$ directly corresponds to real interband transitions, owing to the complex nature of the velocity matrix elements. Using Eq.(\ref{eq:kuboformula}), we obtain the following expression for the imaginary part of transverse optical conductivity as 

\begin{align} \label{eq:fsigmaxy}
\mathrm{Im}(\sigma_{xy})= \frac{e^2}{2\pi} \frac{1}{2 \omega} \int \mathrm{d}k_x \frac{2\alpha k_xd_z'-v_Ld_x'}{\sqrt{{(\omega/2)^2-d_x'^2-d_z'^2}}} ~\Gamma(k_x,\omega,T,\mu)~\Theta\bigg[\bigg(\frac{\omega}{2}\bigg)^2-d_x'^2-d_z'^2\bigg].
\end{align}
 \end{widetext}

\begin{figure}[thb]
\centering
\begin{minipage}[t]{0.5\textwidth}
\hspace{-0.6cm}
\subfigure[]
{\label{fig:tgdrude_xx}\includegraphics[height=5cm,width=0.54\linewidth]{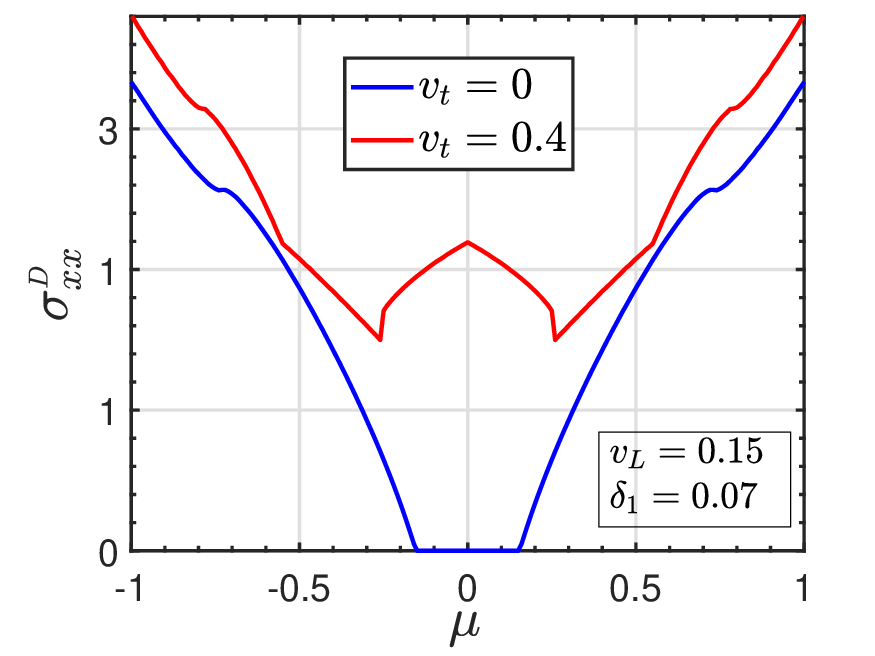}}
\hspace{-0.6cm}
\subfigure[]
{\label{fig:tgdrude_yy}\includegraphics[height=5.2cm,width=0.54\linewidth]{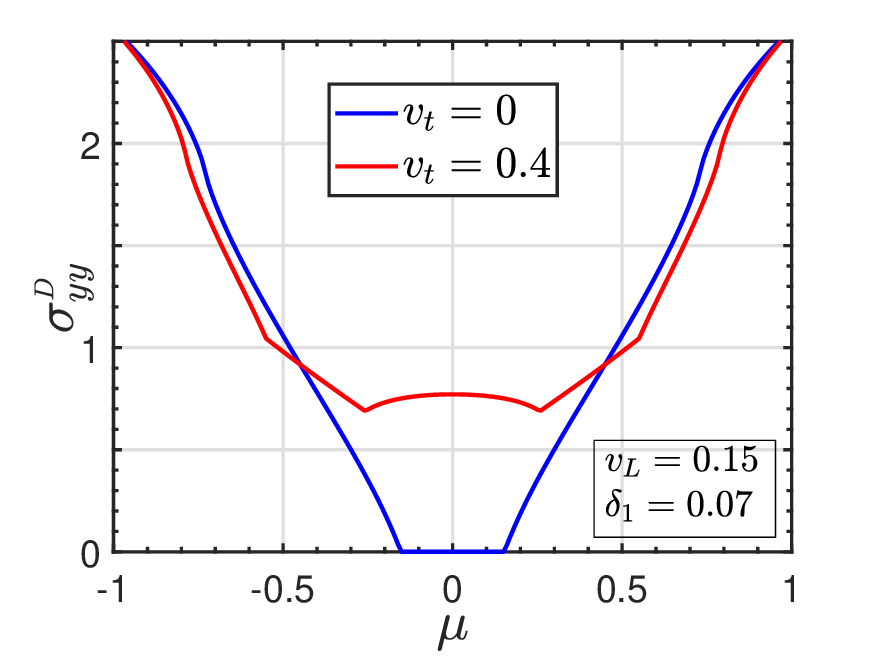}}
\end{minipage}
\caption{Plot of Drude conductivity as a function of chemical potential for non-tilted and tilted SD material irradiated with time-periodic circularly polarized light. Drude conductivity is taken in unit $e^2/2\pi^2$and $\mu$ in the unit of $v_F k_F$.}
\label{fig:irradiated_drude}
\end{figure}
The  two equations, Eqs.~(\ref{eq:fsigmaxx}) and (\ref{eq:fsigmayy}) can be reduced to Eqs.~(\ref{eqn:sigma_xx}) and ~(\ref{eqn:sigma_yy}) by turning off the irradiation, i.e., $\{\delta_1,v_L\}\rightarrow0$. The interband optical conductivities corresponding to Eqs.(\ref{eq:fsigmaxx}),~(\ref{eq:fsigmayy}) and~(\ref{eq:fsigmaxy}) are numerically plotted in Fig.(\ref{fig:irr_optical cond.}). It shows that optical excitation does not take place at the two nodes point until the frequency of applied bias becomes higher than the gap at the nodes. Similar to the non-irradiated case, the optical excitation around the nodes are relatively much stronger in $\mathrm{Re}(\sigma_{xx})$ as shown in Fig.~\ref{fig:tgsigmaxx} in comparison to $\mathrm{Re}(\sigma_{yy})$ on Fig.~\ref{fig:tgsigmayy}. However, the giant peak persists in the $y$-component with restricted energy region at $\omega=2\delta'$ Similarly, the topological gaps are also clearly visible in dynamical Hall conductivity as in Fig~\ref{fig:tgsigmaxy}. The Hall response exhibits sharp peaks at very near to the gap edge, and immediately starts decreasing with the further increase in $\omega$. However, at undoped situation when $\omega\sim 1.5$, Hall response regains some strength because of the high DOS at the band edges, as noted in the $y$-component in \ref{fig:tgsigmayy}.

\subsection{Intraband Drude conductivity of irradiated tilted SD system at $\omega\rightarrow 0$}
 Finally, we revisit the Drude conductivity of irradiated tilted SD system. The two orthogonal components of the Drude conductivity for such a case are obtained as 
\begin{multline}   
\sigma_{xx}^D = \frac{e^2}{2\pi^2v_F}\int \mathrm{d}k_x~\\\times
\frac{\left[v_t \abs{\mu - d_0} +\eta~2\sqrt{\alpha(d_x'+\delta')}d_x'+v_L d_z'\right]^2}
{\abs{\mu - d_0} \sqrt{(\mu - d_0)^2 - d_x'^2-d_z'^2}}\\\times\Theta\left[(\mu-d_0)^2-d_x'^2-d_z'^2\right],
 \end{multline}
 and
\begin{multline}
\sigma_{yy}^D = \frac{e^2 v_F}{2\pi^2} \int \mathrm{d}k_x~\frac{\sqrt{(\mu - d_0)^2 - d_x'^2 - d_z'^2}}{|\mu - d_0|} \\
\Theta\left[(\mu - d_0)^2 - d_x'^2 - d_z'^2\right].
\end{multline}
Which are numerically plotted in Fig.~(\ref{fig:irradiated_drude}). For the non-tilted case ($v_t=0$ and shown by blue), the Drude conductivity completely vanishes as long as $\mu$ lies inside the gap. In contrast to this, for the case of tilted SD system (as shown by red), at the charge neutrality point, the Drude conductivity is finite in both components which is due to tilt induced non-zero velocity and DOS. A slight anomaly between two components is attributed to the difference in group velocities $v_x$ and $v_y$.

\section{Nonlinear Optical response: } \label{sec:nonlinear}
In this section, we discuss the nonlinear optical response of the tilted SD system.  The higher order correction to the Kubo formalism to the external drive can be used to investigate the nonlinear response. We adopt the formalism simplified after performing Matsubara summation, in Refs. [\onlinecite{parker2019,moore2021,omidtavakol2023}]. The second order current density generated in response to external time dependent field can be expressed in terms of nonlinear conductivity tensor $J_{\alpha}(\omega)=\sigma_{\alpha\beta\gamma}(\omega;\omega_1,\omega_2)E_\beta(\omega_1)E_\gamma(\omega_2)$ with $\omega$ as output frequency and $\omega_1$ and $\omega_2$ as frequencies of incident light. We limit our discussion to the SHG, in which the material with broken inversion symmetry is externally driven by a perturbation with frequency $\omega$ and produces optical response at frequency $2\omega$ and $\omega$. We intrinsically break the inversion symmetry by incorporating the constant mass term $\mathrm{m}_z\sigma_z$ to the effective tilted-SD Hamiltonian $\mathcal{H}$.  Such mass term can be included by placing the material on a substrate like hBN \cite{PhysRevB.76.073103}. Therefore, the resulting Hamiltonian is given by $\mathcal{H}_m=d_0\mathbb{1}+\mathbf {d}_{m}\cdot\sigma$ , with $\mathbf{d}_{m}=\{d_x,d_y,m_z\}$. The resulting energy spectrum and energy eigen states are $\mathcal{E}^m_{\eta} = d_0 +\eta\abs{d_m}$
 and $\la {\bf r}|\mathbf{k},\eta\ra= \begin{bmatrix}
 (m_z +\eta\abs{d_m})/(d_x+id_y) & 1 \end{bmatrix}^{T}e^{i\mathbf{k}\cdot r}/N $ respectively, where $N$ is normalization constant. 
The two energy bands are separated by energy gap $2\abs{d_m}$.
The SHG for two-band system in the velocity gauge can be written as follows \cite{moore2021}

\begin{widetext}
\begin{align}
\sigma_{\alpha\beta\gamma}(2\omega;\omega,\omega)=\sigma_{\alpha\beta\gamma}^{2pI}(2\omega;\omega,\omega)+\sigma_{\alpha\beta\gamma}^{2pII}(2\omega;\omega,\omega)+\sigma_{\alpha\beta\gamma}^{1pI}(2\omega;\omega,\omega)+\sigma_{\alpha\beta\gamma}^{1pII}(2\omega;\omega,\omega)
\end{align} 
where,
\begin{align} \label{eqn:2pI}
\sigma^{2pI}_{\alpha\beta\gamma}(2\omega;\omega,\omega)=\frac{i\pi e^3}{2\omega^2}\int\frac{d^2k}{(2\pi)^2}~Y^{2pI}_{\alpha\beta\gamma,+-}\left[f(\mathcal{E}^m_{\bf k,-}) - f(\mathcal{E}^m_{\bf k,+})\right]\delta(2\omega - 2\abs{d_m}
  \end{align}
  \begin{align} \label{eqn:2pII}
\sigma^{2pII}_{\alpha\beta\gamma}(2\omega;\omega,\omega)=\frac{i\pi e^3}{2\omega^2}\int\frac{d^2k}{(2\pi)^2} ~Y^{2pII}_{\alpha\beta\gamma,+-}\left[f(\mathcal{E}^m_{\bf k,-}) - f(\mathcal{E}^m_{\bf k,+})\right]\delta(2\omega - 2\abs{d_m}
  \end{align}

\begin{align}\label{eqn:1pI}
\sigma^{1pI}_{\alpha\beta\gamma}(2\omega;\omega,\omega)=\frac{i\pi e^3}{2\omega^2}\int\frac{d^2k}{(2\pi)^2} ~Y^{1pI}_{\alpha\beta\gamma,+-}\left[f(\mathcal{E}^m_{\bf k,-}) - f(\mathcal{E}^m_{\bf k,+})\right]\delta(\omega - 2\abs{d_m})
\end{align}
\begin{align}\label{eqn:1pII}
\sigma^{1pII}_{\alpha\beta\gamma}(2\omega;\omega,\omega)=\frac{i\pi e^3}{2\omega^2}\int\frac{d^2k}{(2\pi)^2}~Y^{1pII}_{\alpha\beta\gamma,+-}\left[f(\mathcal{E}^m_{\bf k,-}) - f(\mathcal{E}^m_{\bf k,+})\right]\delta(\omega - 2\abs{d_m})
\end{align}

with
\begin{align} \label{eqn:2p1}
Y^{2 p I}_{\alpha\beta\gamma,+-}= v_{\alpha,-+}v_{\beta\gamma,+-}
\end{align}

\begin{align} \label{eqn:2p2}
Y^{2pII}_{\alpha\beta\gamma,+-}=2\frac{v_{\alpha,-+}}{\mathcal{E}^{m}_+-\mathcal{E}^{m}_-}\left[ v_{\beta,+-}\Delta_\gamma\right]_+
\end{align}

\begin{align} \label{eqn:1p1}
Y^{1 p I}_{\alpha\beta\gamma,+-}= v_{\alpha\beta,-+}v_{\gamma,+-}+~v_{\alpha\gamma,-+}v_{\beta,+-}
\end{align}

\begin{align} \label{eqn:1p2}
Y^{1pII}_{\alpha\beta\gamma,+-}=\frac{[v_{\beta,+-}~v_{\gamma,-+}]_+}{2(\mathcal{E}^m_+-\mathcal{E}_-^{m})}\Delta_\alpha~-~ \frac{v_{\alpha,-+}[v_{\beta,+-}~\Delta_\gamma]_+}{(\mathcal{E}^m_+-\mathcal{E}^m_-)}
\end{align}
\end{widetext}
These equations, Eqs.~(\ref{eqn:2pI}~--~\ref{eqn:1pII}) indexed as $2$p and $1$p correspond to the two-photon and one-photon process respectively. Also, $\alpha,\beta,\gamma\in\{x,y\}$, 
\begin{eqnarray}
v_{\alpha,-+}&=&\langle \mathbf{k},-|\frac{\partial\mathcal{H}_{m}}{\partial k_{\alpha}}| \mathbf{k},+\rangle\nonumber\\
v_{\alpha\beta,-+}&=&\langle \mathbf{k},-|\frac{\partial^2\mathcal{H}_{m}}{\partial k_{\alpha}\partial  k_{\beta}}|\mathbf{k},+\rangle
\end{eqnarray}
Again,
\begin{equation}
 [v_{\alpha,-+}v_{\beta,+-}]_{+}=v_{\alpha,-+}v_{\beta,+-}+v_{\beta,-+}v_{\alpha,+-},
\end{equation}
\begin{figure*}[t]
\centering
\subfigure[]
{\label{fig:sigmaxxy2p2} \includegraphics[ height=5cm, width=0.32\linewidth]{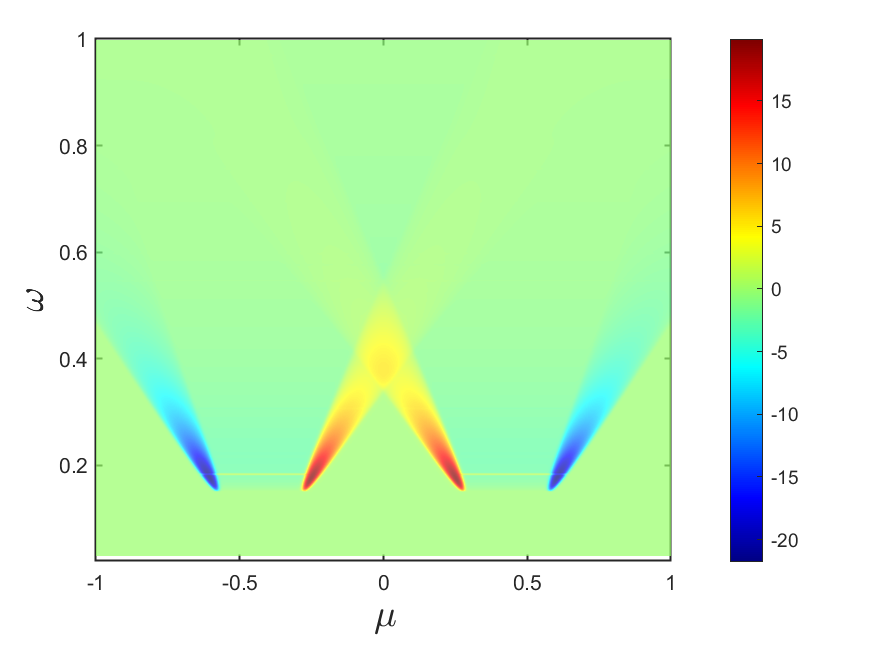}}
\centering
\subfigure[]
{ \label{fig:sigmaxxy1p1}\includegraphics[ height=5cm, width=0.32\linewidth]{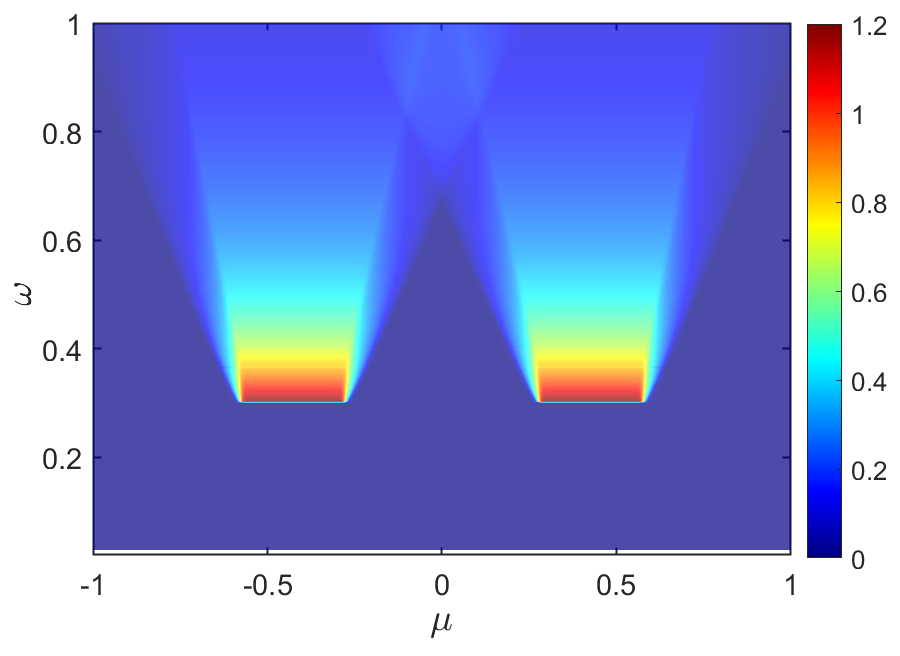}}
\centering
\subfigure[]
{\label{fig:sigmaxxy1p2} \includegraphics[ height=5cm, width=0.32\linewidth]{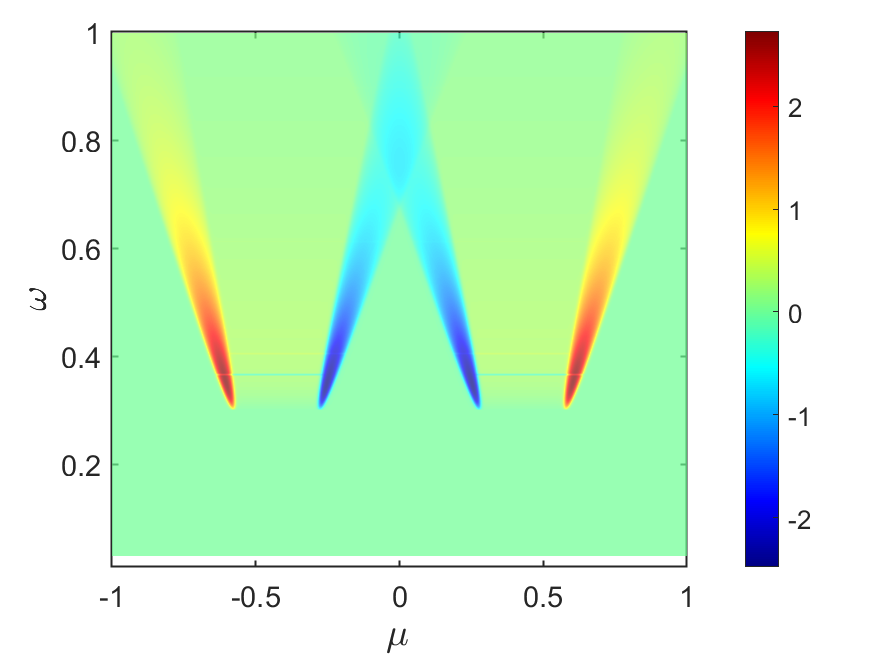}}

\caption{Contour plots showing the real part of the resonant nonlinear optical response(SHG), as a function of photon frequency $\omega$ and chemical potential $\mu$. The color scale represents the magnitude of (a) $\mathrm{Re}[\sigma^{2pII}_{xxy}(\omega, \mu)]$, (b) $\mathrm{Re}[\sigma^{1pI}_{xxy}(\omega, \mu)]$ and (c) $\mathrm{Re}[\sigma^{1pII}_{xxy}(\omega, \mu)]$, in units of $e^3/2\pi v_Fk_F^2$. Here, $m_z=0.15$ in units of $v_F k_F$ .}
\label{fig:nonlinear}
\end{figure*}

  $v_{\alpha,+-}=(v_{\alpha,-+})^*$ and   $\Delta_{\alpha}= v_{\alpha,--}-v_{\alpha,++}$. We now analyze each component of the SHG conductivity tensor. We find that $\operatorname{Re}(\sigma_{xxx})$, $\operatorname{Re}(\sigma_{xyy})$,
$\operatorname{Re}(\sigma_{yyx})$ and $\operatorname{Re}(\sigma_{yyy})$ components of the SHG conductivity vanish as the corresponding ``$Y$'' elements are either real or odd in $k_y$. The only non vanishing components are $\operatorname{Re}(\sigma_{xxy})$ 
and  $\operatorname{Re}(\sigma_{yxx})$  receiving contribution from the imaginary parts of the corresponding components of $Y$ elements. We also note that only the imaginary parts of the  $Y$ elements contribute to the real part of SHG conductivity. The expression for 
$\operatorname{Im}(Y^{2pII}_{xxy})$, $\operatorname{Im}(Y^{1pI}_{xxy})$, 
and $\operatorname{Im}(Y^{1pII}_{xxy})$ are given in Appendix as 
(Eqs.~\eqref{eqn:y1pII}--\eqref{eqn:y1pII}), including the remaining components $\operatorname{Im}(Y^{2pII}_{yxx})$, $\operatorname{Im}(Y^{1pI}_{yyx})$, 
and $\operatorname{Im}(Y^{1pII}_{yxx})$. Therefore, we can write the expressions for ${\rm Re}(\sigma_{xxy})$ as follows
\begin{widetext}
\begin{align}   \label{eq:shg2p2}
\operatorname{Re}(\sigma_{xxy}^{2pII})=\frac{-e^3}{2\pi\omega^4}\int dk_x\frac{(2\alpha k_x)^2m_zd_x}{\sqrt{\omega^2-d_x^2-m^2}}~\Gamma(k_x,2\omega,T,\mu)~\Theta\big[\omega^2-d_x^2 -m_z^2\big]
\end{align}
\begin{align}  \label{eq:shg1p1}
\operatorname{Re}(\sigma_{xxy}^{1pI})=\frac{e^3\alpha m_z}{4\pi\omega^2 }\int \frac{dk_x}{\sqrt{(\omega/2)^2-d_x^2-m_z^2}} ~\Gamma(k_x,\omega,T,\mu)~\Theta\big[(\omega/2)^2-d_x^2-m_z^2 \big]
\end{align}
\begin{align}   \label{eq:shg1p2}
\operatorname{Re}(\sigma_{xxy}^{1pII})=\frac{e^3}{2\pi\omega^4}\int dk_x\frac{2(2\alpha k_x)^2m_zd_x}{\sqrt{(\omega/2)^2-d_x^2-m_z^2}}~\Gamma(k_x,\omega,T,\mu)~\Theta\big[(\omega/2)^2-d_x^2-m_z^2 \big]
\end{align} 

\end {widetext}

 The other components corresponding to the $\operatorname{Re}(\sigma_{yxx})$ differ from the $\operatorname {Re}(\sigma_{xxy})$ only in terms of prefactor as their corresponding $Y$ elements are related ( See Appendix (\ref{sec:appendix})).\\
The  SHG plots corresponding to Eqs. (\ref{eq:shg2p2}), (\ref{eq:shg1p1}) and (\ref{eq:shg1p2})  are numerically plotted in Fig.~(\ref{fig:nonlinear}).  The Fig.~\ref{fig:sigmaxxy2p2} corresponding to the $\operatorname{Re}(\sigma_{xxy}^{2PII})$ shows a sharp resonance  for the threshold frequency $\omega=m_z$ (two-photon resonance) with peak position shifting the sign across the $\mu$ corresponding to the two nodes separated by the gap $2m_z$ on the positive side.  However, the two photon process is strong only in a very narrow regime with minimum value of frequency $\omega\ge(\mu-\mu_c)$ where $\mu_c=\pm v_t\sqrt{\delta_0/\alpha}$ is the position of nodes in energy space. On the other hand, the single photon process consists two contributions as $\operatorname{Re}(\sigma_{xxy}^{1PI})$ and $\operatorname{Re}(\sigma_{xxy}^{1PII})$, which are plotted in Fig.~\ref{fig:sigmaxxy1p1} and Fig.~\ref{fig:sigmaxxy1p2}, respectively. We note that the second contribution, in Fig.~\ref{fig:sigmaxxy1p2}, exhibits almost the same pattern as in two photon process, but the minimum frequency required to initiate the process is $\omega\ge 2(\mu-\mu_c)$. The 1st contribution for single photon process, in the Fig.~\ref{fig:sigmaxxy1p1}, is very strong as long as $\omega$ just overcomes the gap while $\mu$ is kept inside the gap. Moreover, the 1st contribution to the single photon process survives in a relatively broad parameter range.

The $2$ factor difference between the two and one photon process can be also clearly understood from the step function in the above three Eqs~[\ref{eq:shg2p2}-\ref{eq:shg1p2}]. Note that the two photon and one photon processes also differ by opposite phase, which emerges intrinsically from the velocity matrices.
\section{\bf Conclusion}  \label{sec:conclusion}
We have studied the electronic band structure of a tilted SD material with tilt along the quadratic direction that gives rise to the energy-imbalanced nodes. We obtained the analytical expression for the interband anisotropic optical conductivity. We reveal that the tilt induced energy imbalance between two nodes can be probed by optical excitation by looking into its profile in $(\omega,\mu)$ space, particularly the difference between two chemical potential values where excitation starts at near zero frequency is the direct measure of energy imbalance. It has also been noted that even though light opens up topological gap at the two nodes, a system can never obtain bulk insulating phase because of energy imbalance. Such phase can never be probed by anomalous Hall conductivity as it will be suppressed by the bulk contribution. In this case, optical conductivity can be much more relevant than the Hall conductivity, and the minimum frequency required to excite interband optical transition measures the topological gap. We have also studied the DC conductivity using semi-classical Boltzmann transport theory. We revealed that tilt can convert the SD system from semimetallic phase to metallic phase, contrary to the tilted Dirac system. Furthermore, we also studied the SHG of the tilted SD system with intrinsically broken inversion symmetry and observed that the two-photon resonance contribution dominates near threshold frequency, with significantly higher peak intensity in comparison to the one-photon contribution. 
\paragraph{Note:} While revising the manuscript we came across the similar study but without two Weyl nodes in Ref.~[\onlinecite{chen2025arxiv}].\section*{Acknowledgment\label{Sec:acknowledgements}}
The authors thank Tarun  Kanti Ghosh for useful discussions and valuable comments. SK Firoz Islam acknowledges the project ANRF/ECRG/2024/005166/PMS.
\section{Data availability}
The data that supports the findings of this article are not publicly available. The data are available from the authors upon reasonable request.

\begin{figure*}[h]
 \centering
 \subfigure[]
 {\includegraphics[width=0.4\linewidth]{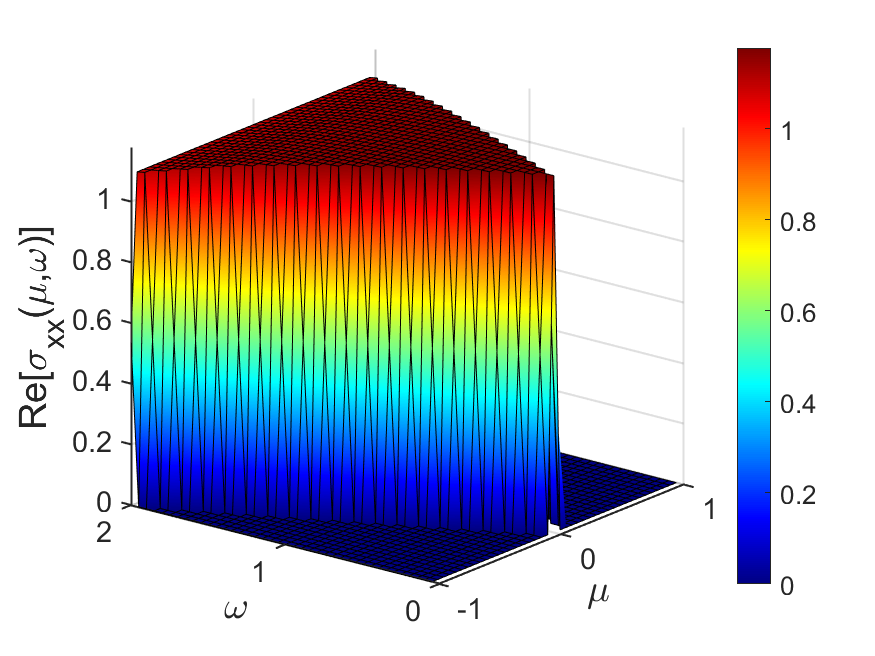}   }
\subfigure[]
{\includegraphics[width=0.4\linewidth]{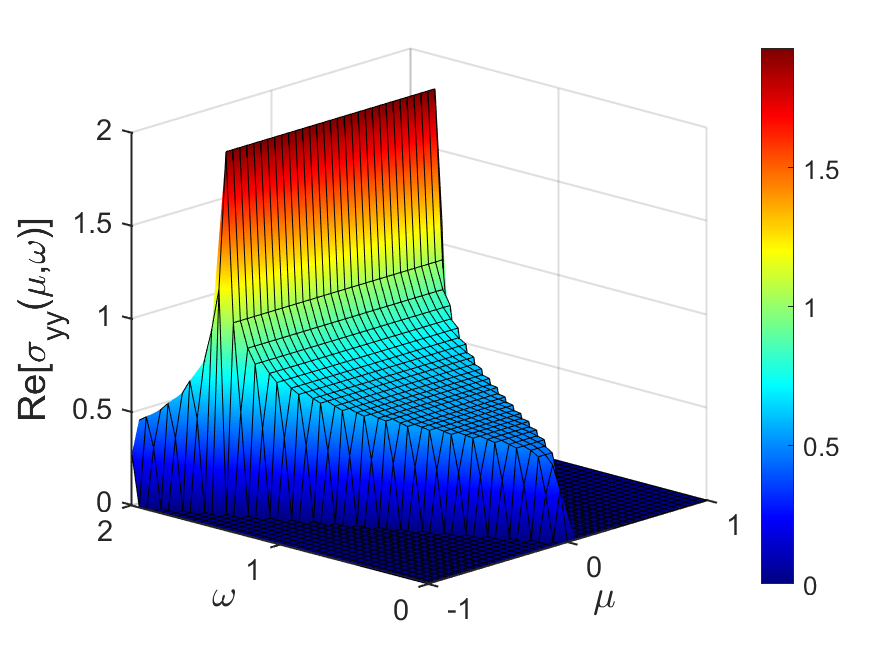} }
\subfigure[]
{\includegraphics[width=0.4\linewidth]{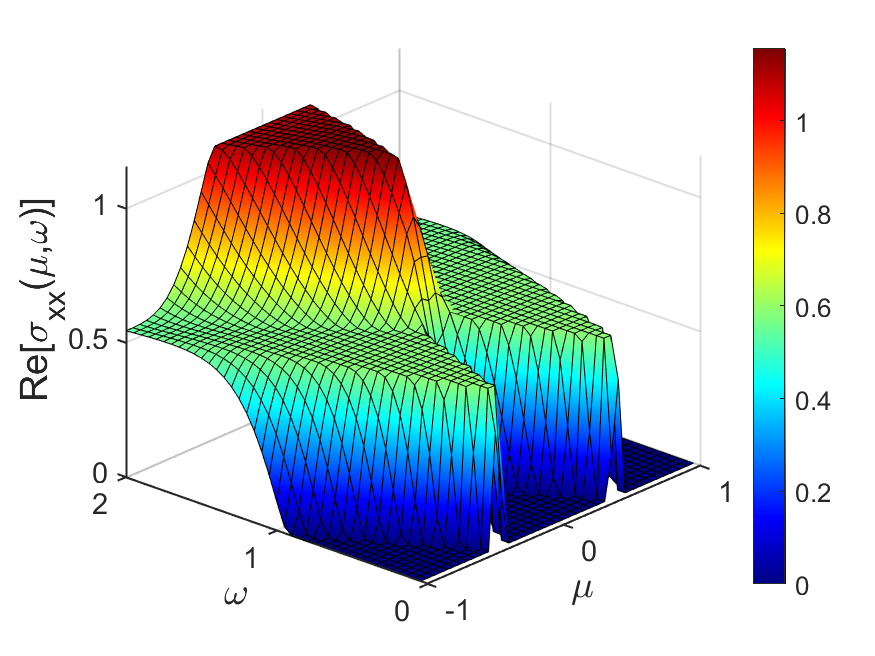} }
\subfigure[]
{\includegraphics[width=0.4\linewidth]{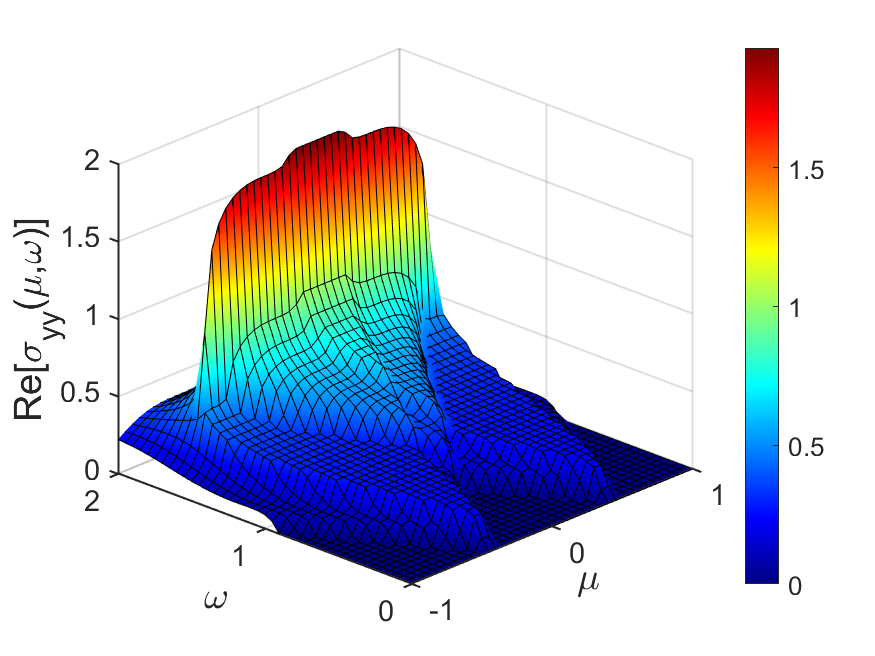} }
\caption{$3$D surface plots of the interband optical conductivity of tilted SD material }
 \label{fig:3dplotsnonirradiated} 
 \end{figure*}


\begin{figure*}[h]
\centering
\subfigure[]{\includegraphics[width=0.32\linewidth]{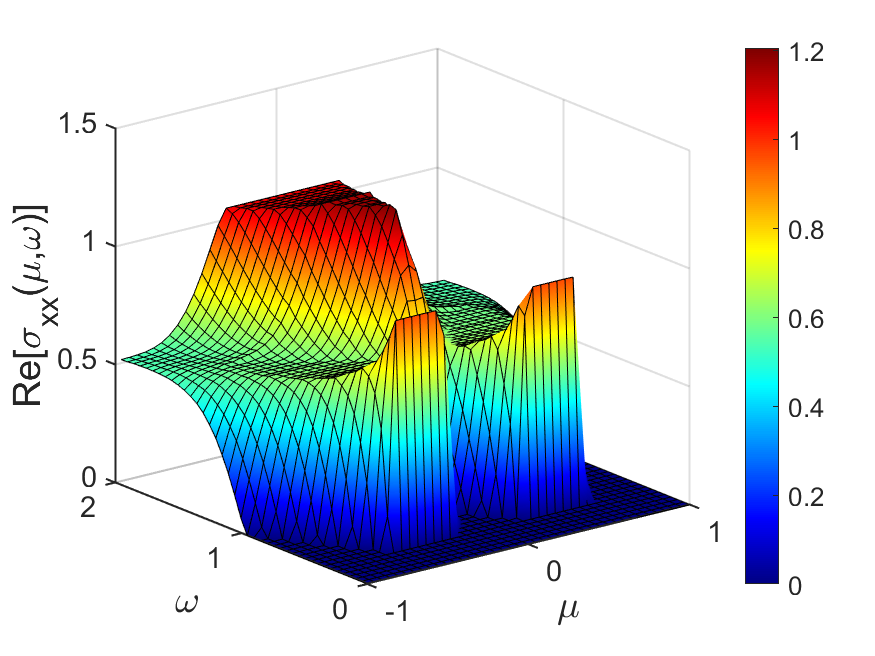}}
\subfigure[]{\includegraphics[width=0.32\linewidth]{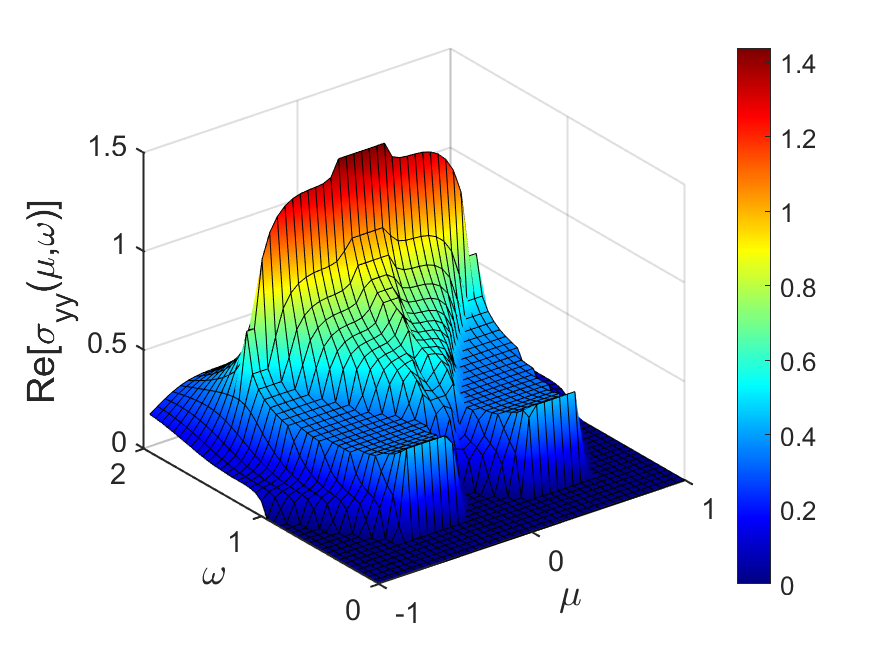}}
\subfigure[]{\includegraphics[width=0.32\linewidth]{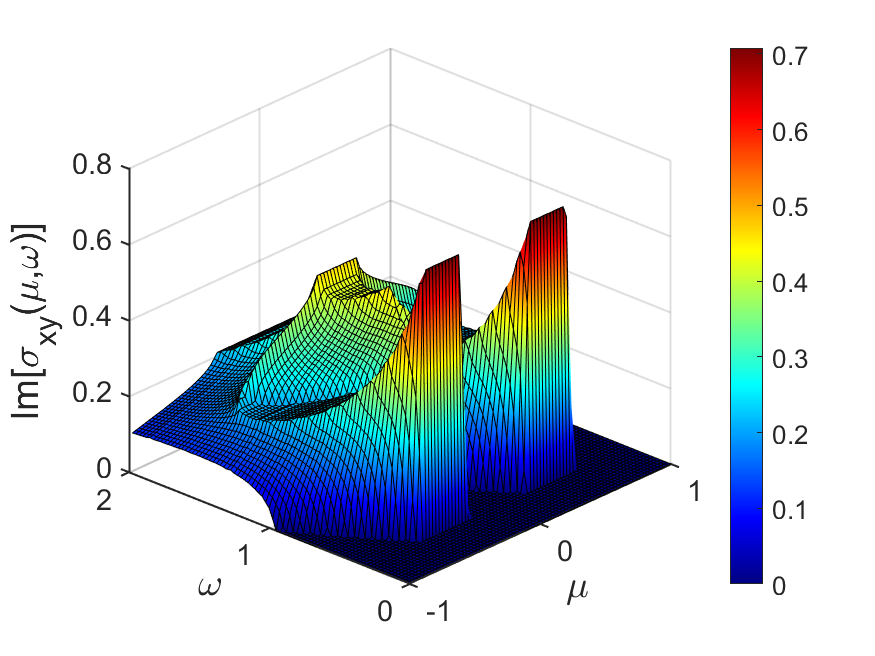}}
\caption{$3$D surface plots of interband optical conductivity for an irradiated tilted SD material.}
\label{fig:3dplotirradiated}
\end{figure*}
\appendix
\section{$3$D plots} \label{sec:3Dplots}
Here we present the three dimensional ($3$D) plots for the real part of optical conductivity using Eq.~(\ref{eqn:sigma_xx}) and Eq.~(\ref{eqn:sigma_yy}) with and without tilt. These plots add  further insight into the variation of the optical conductivity as a function of the chemical potential $\mu$ and frequency $\omega$. In  Fig~\ref{fig:3dplotsnonirradiated} (a) and (b) the Re[$\sigma_{xx}]$ and Re[$\sigma_{yy}$] are plotted, which shows near zero frequency optical excitation starts at $\mu=0$. However, with the increase of frequency the y-component of optical excitation sharply increases around the Van-Hove singularity in DOS at the band edge as shown in DOS plot in Fig.~\ref{fig:dos}. The x-component increases linearly in $\omega$ and $\mu$ both. Once the tilt is switch on, the two Weyl nodes displace itself at two different energy levels at $\mu_c=\pm v_t\sqrt{\delta_0/\alpha}$. Hence, we can see the near zero frequency optical excitation starts at two locations of $\mu$ around $\pm\mu_c$ indicating the clear signature of energy imbalanced nodes Fig.~\ref{fig:3dplotsnonirradiated} (c) and (d). The signature of two individual nodes at different energy is much stronger in x-component than $y$-component.\\
The $3$D plots corresponding to Eqs.~[\ref{eq:fsigmaxx}-\ref{eq:fsigmaxy}] are shown in  Fig.~\ref{fig:3dplotirradiated}.  In Fig.\ref{fig:3dplotirradiated} (a) and (b), the Re[$\sigma_{xx}$] and Re[$\sigma_{yy}$] are plotted for an irradiated tilted SD system.  It is evident that the optical excitation doesn't take place untill the frequency of the applied bias becomes more than the nodal gap. The  Fig.\ref{fig:3dplotirradiated} (c) corresponds to the Hall response, which also shows the visible gap signature and reflects the broken time reversal symmetry induced by circularly polarised irradiation.

\section {Power-law dependence of Optical conductivity of tilted SD system}
\label{power_law}
Here, we try to obtain an approximate analytical form of the interband optical conductivity, aiming to find the power law for $\omega$. We follow the approach as in  Ref.~[\onlinecite{xu_yan2023}]. In the limit $T\rightarrow0$, the Eq.(\ref{eqn:sigma_xx}), can be written as
\begin{widetext}
\begin{align}
\operatorname{Re}(\sigma_{xx})=\frac{2e^2}{\pi} \frac{1}{v_F\omega^2} \int_{-\infty}^{\infty} \mathrm{d}{k_x}~ 
        \alpha^2 k_x^2 \sqrt{\bigg(\frac{\omega}{2}\bigg)^2 - (\alpha k_x^2-\delta_0)^2}
        \times\Theta\bigg[\bigg(\frac{\omega}{2}\bigg)^2-(\alpha k_x^2 -\delta_0)^2 \bigg] \nonumber \\
        \left[\Theta(\mu+\omega/2-v_tk_x)-\Theta(\mu-\omega/2-v_tk_x)\right].
\end{align}
By splitting the $\int_{-\infty}^{\infty}...=\int_{-\infty}^{0}...+\int_{0}^{\infty}...$,and  changing the variable from $k_x$ to $-k_x$ in first, we get 
\begin{align}
\operatorname{Re}(\sigma_{xx})=\frac{2e^2}{\pi} \frac{1}{v_F\omega^2} \int_{0}^{\infty} \mathrm{d}{k_x}~ 
        \alpha^2 k_x^2 \sqrt{\bigg(\frac{\omega}{2}\bigg)^2 - (\alpha k_x^2-\delta_0)^2}
        ~~\Theta\bigg[\bigg(\frac{\omega}{2}\bigg)^2-(\alpha k_x^2 -\delta_0)^2 \bigg] \nonumber \\
        \times\sum_{s=\pm}s\Theta(\mu+s\omega/2-v_tk_x)
\end{align}
Substituting, $\phi=2(\alpha k_x^2-\delta_0)/\omega$,  we have $\mathrm{d} k_x=\omega\mathrm{d}\phi/4\alpha k_x$. Also, $k_x=\sqrt{(\phi\omega+2\delta)/2\alpha}$. Therefore, the equation becomes,
\begin{align} \label{eqn:Sigmaxx}
\operatorname{Re}(\sigma_{xx})=\frac{2e^2}{\pi} \frac{1}{8v_F} \int_{-2\delta_0/\omega}^{\infty} \mathrm{d}{\phi}~ 
         \sqrt{\alpha\delta_0\bigg(1+\frac{\phi\omega}{2\delta_0}\bigg) }
        \times\sqrt{1-\phi^2}\times \mathcal{Q}(v_t,\alpha,\delta_0,\phi,\omega,\mu)
\end{align}
with acceptable range of limit, $\phi=[-1,1]$ and $\mathcal{Q}(v_t,\alpha,\delta_0,\phi,\omega,\mu)$ defined as,
\begin{align}
\mathcal{Q}(v_t,\alpha,\delta_0,\phi,\omega,\mu)=\sum_{s=\pm}s\bigg[\Theta(\mu+s\omega/2-v_t\sqrt{\delta_0/\alpha}\sqrt{(1+\phi\omega/2\delta_0})\bigg]
\end{align}
This $\mathcal{Q}(v_t,\alpha,\delta_0,\phi,\omega,\mu)$ only controls the range of integration.
\paragraph{Case I:} In the limit  $2\delta_0<<\omega$ or $\delta_o\rightarrow0$, the omega dependency can be directly extracted from Eq.~(\ref{eqn:Sigmaxx}) as ,
\begin{align}
\operatorname {Re}(\sigma_{xx})=\sqrt{\omega}\times I_{xx} 
\end{align}
with $I_{xx}= 2 e^2/8\pi v_F \times \sqrt{\alpha/2} \int _{0}^{1}\sqrt{\phi}\sqrt{1-\phi^2}
\times\mathcal{Q}(v_t,\alpha,\delta_0,\phi, \omega ,\mu)$ is the integral free from any $\omega $ dependence.
\paragraph{Case II}: In the limit $2\delta_{0}>>\omega$ , we can see from the Eq.~(\ref{eqn:Sigmaxx}), $\operatorname{Re}(\sigma_{xx})$ will have no $\omega$ dependence.
\paragraph{Case III:} In the limit $2\delta_{0}\sim\omega$, no clear explict $\omega$ dependence can be extracted.
\vspace{1cm}
\par In the similar, we can write   $\operatorname{Re}(\sigma_{yy})$ as,
\begin{align} \label{eqn:Sigmayy}
\operatorname{Re}(\sigma_{yy})=\frac{e^2v_F}{2\pi} \frac{1}{8\alpha} \int_{-2\delta_0/\omega}^{\infty} \mathrm{d}{\phi}~ 
         \frac{1}{\sqrt{\alpha\delta_0\bigg(1+\frac{\phi\omega}{2\delta_0}\bigg)} }
        \times\frac{\phi^2}{\sqrt{1-\phi^2}}\times \mathcal{Q}(v_t,\alpha,\delta_0,\phi,\omega,\mu)
\end{align}
with range of limit, $\phi=[-1,1]$
\paragraph{Case I:} In the limit  $2\delta_0<<\omega$ or $\delta_o\rightarrow0$, the omega dependency can be directly extracted from Eq.~(\ref{eqn:Sigmayy}) as ,
\begin{align}
\operatorname {Re}(\sigma_{yy})=\frac{1}{\sqrt{\omega}}\times I_{yy} 
\end{align}
with $I_{yy}= e^2 v_{F}/8\sqrt{2}\pi\alpha^{3/2}\times\int_{0}^{1} \phi^{3/2}\times(1-\phi^2)^{-1/2}
\times \mathcal{Q}(v_t,\alpha,\delta_0,\phi,\omega,\mu)$ as the integral free from any $\omega $ dependence.
\paragraph{Case II:} In the limit $2\delta_{0}>>\omega$ , we can see from the Eq.~(\ref{eqn:Sigmayy}), $\operatorname{Re}(\sigma_{yy})$ will have no $\omega$ dependence.
\paragraph{Case III:} In the limit $2\delta_{0}\sim\omega$, no clear explicit $\omega$ dependence can be extracted.
\end{widetext}

\section{Matrix elements }\label{sec:appendix}
 The different matrix elements $v_\alpha$, $v_{\alpha\beta}$ and $\Delta_\alpha$ for different $x,y$ directions are given as
 \begin{align*}
v_{x,+-}=2\alpha k_x\frac{d_xm_z+id_y\abs{d_m}}{|d_m|\abs{d}}
;\quad
  v_{xx,+-}=2\alpha \frac{d_xm_z+id_y\abs{d_m}}{|d_m|\abs{d}}
\end{align*}
 \begin{align*}
 v_{y,+-}=v_F\frac{d_ym_z-id_x\abs{d_m}}{|d_m|\abs{d}}\end{align*} 
\begin{align*}
v_{x,\pm\pm}=v_t\pm\frac{2(2\alpha k_x)d_x}{\abs{d_m}}
\quad;
v_{y,\pm\pm}=\pm\frac{2v_Fd_y}{\abs{d_m}}
\end{align*}
Also, the components $v_{yy,+-}$,$v_{xy,+-}$ vanish as the Hamiltonian($\mathcal{H}_{m}$) is linear along the $y$ direction and no cross term $xy$ exists . Using the above matrix elements in Eqs.(\ref{eqn:2p1}--\ref{eqn:1p2}), we calculate all the components $Y^{2 p I}_{\alpha\beta\gamma,+-}$,$Y^{2 p II}_{\alpha\beta\gamma,+-}$,$Y^{1 p I}_{\alpha\beta\gamma,+-}$ and $Y^{1 p II}_{\alpha\beta\gamma,+-}$ for all combinations of $\alpha,\beta,\gamma=(x,y)$. 
The different components of $\operatorname{Im}(Y_{\alpha\beta\gamma,+-})$  that give rise to the real part of SHG are found to be
\begin{align} \label{eqn:y2pII}
\operatorname{Im}(Y_{xxy,+-}^{2 p II})=\frac{4v_F(2\alpha k_x)^2m_zd_x}{(\abs{d_m})^3}
\end{align}
\begin{align}  \label{eqn:y1pI}
\operatorname{Im}(Y^{xxy}_{1 p I})=-\frac{2\alpha v_Fm_z}{(\abs{d_m})}
\end{align}
\begin{align}  \label{eqn:y1pII}
\operatorname{Im}(Y^{xxy}_{1 p II})=-\frac{2(2\alpha k_x)^2 v_Fd_xm_z}{(\abs{d_m})^3}
\end{align}
The other components $\operatorname{Im}(Y_{yxx,+-}^{2pI})=-\operatorname{Im}(Y_{xxy,+-}^{1 p I})$, $\operatorname{Im}(Y_{yxx,+-}^{2pII})=-2\operatorname{Im}(Y_{xxy,+-}^{2 p II})$ and $\operatorname{Im}(Y_{yxx,+-}^{1pII})=-2\operatorname{Im}(Y_{xxy,+-}^{1 p II})$.

\bibliography{Sd}

\begin{thebibliography}{64}%
\makeatletter
\providecommand \@ifxundefined [1]{%
 \@ifx{#1\undefined}
}%
\providecommand \@ifnum [1]{%
 \ifnum #1\expandafter \@firstoftwo
 \else \expandafter \@secondoftwo
 \fi
}%
\providecommand \@ifx [1]{%
 \ifx #1\expandafter \@firstoftwo
 \else \expandafter \@secondoftwo
 \fi
}%
\providecommand \natexlab [1]{#1}%
\providecommand \enquote  [1]{``#1''}%
\providecommand \bibnamefont  [1]{#1}%
\providecommand \bibfnamefont [1]{#1}%
\providecommand \citenamefont [1]{#1}%
\providecommand \href@noop [0]{\@secondoftwo}%
\providecommand \href [0]{\begingroup \@sanitize@url \@href}%
\providecommand \@href[1]{\@@startlink{#1}\@@href}%
\providecommand \@@href[1]{\endgroup#1\@@endlink}%
\providecommand \@sanitize@url [0]{\catcode `\\12\catcode `\$12\catcode
  `\&12\catcode `\#12\catcode `\^12\catcode `\_12\catcode `\%12\relax}%
\providecommand \@@startlink[1]{}%
\providecommand \@@endlink[0]{}%
\providecommand \url  [0]{\begingroup\@sanitize@url \@url }%
\providecommand \@url [1]{\endgroup\@href {#1}{\urlprefix }}%
\providecommand \urlprefix  [0]{URL }%
\providecommand \Eprint [0]{\href }%
\providecommand \doibase [0]{https://doi.org/}%
\providecommand \selectlanguage [0]{\@gobble}%
\providecommand \bibinfo  [0]{\@secondoftwo}%
\providecommand \bibfield  [0]{\@secondoftwo}%
\providecommand \translation [1]{[#1]}%
\providecommand \BibitemOpen [0]{}%
\providecommand \bibitemStop [0]{}%
\providecommand \bibitemNoStop [0]{.\EOS\space}%
\providecommand \EOS [0]{\spacefactor3000\relax}%
\providecommand \BibitemShut  [1]{\csname bibitem#1\endcsname}%
\let\auto@bib@innerbib\@empty
\bibitem [{\citenamefont {Dietl}\ \emph {et~al.}(2008)\citenamefont {Dietl},
  \citenamefont {Pi\'echon},\ and\ \citenamefont {Montambaux}}]{Dietl2008}%
  \BibitemOpen
  \bibfield  {author} {\bibinfo {author} {\bibfnamefont {P.}~\bibnamefont
  {Dietl}}, \bibinfo {author} {\bibfnamefont {F.}~\bibnamefont {Pi\'echon}},\
  and\ \bibinfo {author} {\bibfnamefont {G.}~\bibnamefont {Montambaux}},\
  }\bibfield  {title} {\bibinfo {title} {New magnetic field dependence of
  landau levels in a graphenelike structure},\ }\href
  {https://doi.org/10.1103/PhysRevLett.100.236405} {\bibfield  {journal}
  {\bibinfo  {journal} {Phys. Rev. Lett.}\ }\textbf {\bibinfo {volume} {100}},\
  \bibinfo {pages} {236405} (\bibinfo {year} {2008})}\BibitemShut {NoStop}%
\bibitem [{\citenamefont {Banerjee}\ \emph {et~al.}(2009)\citenamefont
  {Banerjee}, \citenamefont {Singh}, \citenamefont {Pardo},\ and\ \citenamefont
  {Pickett}}]{Banerjee_pickett2009}%
  \BibitemOpen
  \bibfield  {author} {\bibinfo {author} {\bibfnamefont {S.}~\bibnamefont
  {Banerjee}}, \bibinfo {author} {\bibfnamefont {R.~R.~P.}\ \bibnamefont
  {Singh}}, \bibinfo {author} {\bibfnamefont {V.}~\bibnamefont {Pardo}},\ and\
  \bibinfo {author} {\bibfnamefont {W.~E.}\ \bibnamefont {Pickett}},\
  }\bibfield  {title} {\bibinfo {title} {Tight-binding modeling and low-energy
  behavior of the semi-dirac point},\ }\href
  {https://doi.org/10.1103/PhysRevLett.103.016402} {\bibfield  {journal}
  {\bibinfo  {journal} {Phys. Rev. Lett.}\ }\textbf {\bibinfo {volume} {103}},\
  \bibinfo {pages} {016402} (\bibinfo {year} {2009})}\BibitemShut {NoStop}%
\bibitem [{\citenamefont {Saha}(2016)}]{kush2016}%
  \BibitemOpen
  \bibfield  {author} {\bibinfo {author} {\bibfnamefont {K.}~\bibnamefont
  {Saha}},\ }\bibfield  {title} {\bibinfo {title} {Photoinduced chern
  insulating states in semi-dirac materials},\ }\href
  {https://doi.org/10.1103/PhysRevB.94.081103} {\bibfield  {journal} {\bibinfo
  {journal} {Phys. Rev. B}\ }\textbf {\bibinfo {volume} {94}},\ \bibinfo
  {pages} {081103} (\bibinfo {year} {2016})}\BibitemShut {NoStop}%
\bibitem [{\citenamefont {Shao}\ \emph {et~al.}(2024)\citenamefont {Shao},
  \citenamefont {Moon}, \citenamefont {Rudenko}, \citenamefont {Wang},
  \citenamefont {Herzog-Arbeitman}, \citenamefont {Ozerov}, \citenamefont
  {Graf}, \citenamefont {Sun}, \citenamefont {Queiroz}, \citenamefont {Lee},
  \citenamefont {Zhu}, \citenamefont {Mao}, \citenamefont {Katsnelson},
  \citenamefont {Bernevig}, \citenamefont {Smirnov}, \citenamefont {Millis},\
  and\ \citenamefont {Basov}}]{PhysRevX.14.041057}%
  \BibitemOpen
  \bibfield  {author} {\bibinfo {author} {\bibfnamefont {Y.}~\bibnamefont
  {Shao}}, \bibinfo {author} {\bibfnamefont {S.}~\bibnamefont {Moon}}, \bibinfo
  {author} {\bibfnamefont {A.~N.}\ \bibnamefont {Rudenko}}, \bibinfo {author}
  {\bibfnamefont {J.}~\bibnamefont {Wang}}, \bibinfo {author} {\bibfnamefont
  {J.}~\bibnamefont {Herzog-Arbeitman}}, \bibinfo {author} {\bibfnamefont
  {M.}~\bibnamefont {Ozerov}}, \bibinfo {author} {\bibfnamefont
  {D.}~\bibnamefont {Graf}}, \bibinfo {author} {\bibfnamefont {Z.}~\bibnamefont
  {Sun}}, \bibinfo {author} {\bibfnamefont {R.}~\bibnamefont {Queiroz}},
  \bibinfo {author} {\bibfnamefont {S.~H.}\ \bibnamefont {Lee}}, \bibinfo
  {author} {\bibfnamefont {Y.}~\bibnamefont {Zhu}}, \bibinfo {author}
  {\bibfnamefont {Z.}~\bibnamefont {Mao}}, \bibinfo {author} {\bibfnamefont
  {M.~I.}\ \bibnamefont {Katsnelson}}, \bibinfo {author} {\bibfnamefont
  {B.~A.}\ \bibnamefont {Bernevig}}, \bibinfo {author} {\bibfnamefont
  {D.}~\bibnamefont {Smirnov}}, \bibinfo {author} {\bibfnamefont {A.~J.}\
  \bibnamefont {Millis}},\ and\ \bibinfo {author} {\bibfnamefont {D.~N.}\
  \bibnamefont {Basov}},\ }\bibfield  {title} {\bibinfo {title} {Semi-dirac
  fermions in a topological metal},\ }\href
  {https://doi.org/10.1103/PhysRevX.14.041057} {\bibfield  {journal} {\bibinfo
  {journal} {Phys. Rev. X}\ }\textbf {\bibinfo {volume} {14}},\ \bibinfo
  {pages} {041057} (\bibinfo {year} {2024})}\BibitemShut {NoStop}%
\bibitem [{\citenamefont {Mandal}\ and\ \citenamefont
  {Saha}(2020)}]{IpshitaKush}%
  \BibitemOpen
  \bibfield  {author} {\bibinfo {author} {\bibfnamefont {I.}~\bibnamefont
  {Mandal}}\ and\ \bibinfo {author} {\bibfnamefont {K.}~\bibnamefont {Saha}},\
  }\bibfield  {title} {\bibinfo {title} {Thermopower in an anisotropic
  two-dimensional weyl semimetal},\ }\href
  {https://doi.org/10.1103/PhysRevB.101.045101} {\bibfield  {journal} {\bibinfo
   {journal} {Phys. Rev. B}\ }\textbf {\bibinfo {volume} {101}},\ \bibinfo
  {pages} {045101} (\bibinfo {year} {2020})}\BibitemShut {NoStop}%
\bibitem [{\citenamefont {Pardo}\ and\ \citenamefont
  {Pickett}(2009)}]{pardo_pickett2009}%
  \BibitemOpen
  \bibfield  {author} {\bibinfo {author} {\bibfnamefont {V.}~\bibnamefont
  {Pardo}}\ and\ \bibinfo {author} {\bibfnamefont {W.~E.}\ \bibnamefont
  {Pickett}},\ }\bibfield  {title} {\bibinfo {title} {Half-metallic
  semi-dirac-point generated by quantum confinement in
  ${\mathrm{tio}}_{2}/{\mathrm{vo}}_{2}$ nanostructures},\ }\href
  {https://doi.org/10.1103/PhysRevLett.102.166803} {\bibfield  {journal}
  {\bibinfo  {journal} {Phys. Rev. Lett.}\ }\textbf {\bibinfo {volume} {102}},\
  \bibinfo {pages} {166803} (\bibinfo {year} {2009})}\BibitemShut {NoStop}%
\bibitem [{\citenamefont {Montambaux}\ \emph {et~al.}(2009)\citenamefont
  {Montambaux}, \citenamefont {Pi\'echon}, \citenamefont {Fuchs},\ and\
  \citenamefont {Goerbig}}]{Montambaux2009}%
  \BibitemOpen
  \bibfield  {author} {\bibinfo {author} {\bibfnamefont {G.}~\bibnamefont
  {Montambaux}}, \bibinfo {author} {\bibfnamefont {F.}~\bibnamefont
  {Pi\'echon}}, \bibinfo {author} {\bibfnamefont {J.-N.}\ \bibnamefont
  {Fuchs}},\ and\ \bibinfo {author} {\bibfnamefont {M.~O.}\ \bibnamefont
  {Goerbig}},\ }\bibfield  {title} {\bibinfo {title} {Merging of dirac points
  in a two-dimensional crystal},\ }\href
  {https://doi.org/10.1103/PhysRevB.80.153412} {\bibfield  {journal} {\bibinfo
  {journal} {Phys. Rev. B}\ }\textbf {\bibinfo {volume} {80}},\ \bibinfo
  {pages} {153412} (\bibinfo {year} {2009})}\BibitemShut {NoStop}%
\bibitem [{\citenamefont {Delplace}\ and\ \citenamefont
  {Montambaux}(2010)}]{delpace}%
  \BibitemOpen
  \bibfield  {author} {\bibinfo {author} {\bibfnamefont {P.}~\bibnamefont
  {Delplace}}\ and\ \bibinfo {author} {\bibfnamefont {G.}~\bibnamefont
  {Montambaux}},\ }\bibfield  {title} {\bibinfo {title} {Semi-dirac point in
  the hofstadter spectrum},\ }\href
  {https://doi.org/10.1103/PhysRevB.82.035438} {\bibfield  {journal} {\bibinfo
  {journal} {Phys. Rev. B}\ }\textbf {\bibinfo {volume} {82}},\ \bibinfo
  {pages} {035438} (\bibinfo {year} {2010})}\BibitemShut {NoStop}%
\bibitem [{\citenamefont {Adroguer}\ \emph {et~al.}(2016)\citenamefont
  {Adroguer}, \citenamefont {Carpentier}, \citenamefont {Montambaux},\ and\
  \citenamefont {Orignac}}]{Adroguer2016}%
  \BibitemOpen
  \bibfield  {author} {\bibinfo {author} {\bibfnamefont {P.}~\bibnamefont
  {Adroguer}}, \bibinfo {author} {\bibfnamefont {D.}~\bibnamefont
  {Carpentier}}, \bibinfo {author} {\bibfnamefont {G.}~\bibnamefont
  {Montambaux}},\ and\ \bibinfo {author} {\bibfnamefont {E.}~\bibnamefont
  {Orignac}},\ }\bibfield  {title} {\bibinfo {title} {Diffusion of dirac
  fermions across a topological merging transition in two dimensions},\ }\href
  {https://doi.org/10.1103/PhysRevB.93.125113} {\bibfield  {journal} {\bibinfo
  {journal} {Phys. Rev. B}\ }\textbf {\bibinfo {volume} {93}},\ \bibinfo
  {pages} {125113} (\bibinfo {year} {2016})}\BibitemShut {NoStop}%
\bibitem [{\citenamefont {Zhou}\ \emph {et~al.}(2021)\citenamefont {Zhou},
  \citenamefont {Chen},\ and\ \citenamefont {Zhu}}]{zhouchen2021}%
  \BibitemOpen
  \bibfield  {author} {\bibinfo {author} {\bibfnamefont {X.}~\bibnamefont
  {Zhou}}, \bibinfo {author} {\bibfnamefont {W.}~\bibnamefont {Chen}},\ and\
  \bibinfo {author} {\bibfnamefont {X.}~\bibnamefont {Zhu}},\ }\bibfield
  {title} {\bibinfo {title} {Anisotropic magneto-optical absorption and linear
  dichroism in two-dimensional semi-dirac electron systems},\ }\href
  {https://doi.org/10.1103/PhysRevB.104.235403} {\bibfield  {journal} {\bibinfo
   {journal} {Phys. Rev. B}\ }\textbf {\bibinfo {volume} {104}},\ \bibinfo
  {pages} {235403} (\bibinfo {year} {2021})}\BibitemShut {NoStop}%
\bibitem [{\citenamefont {Sinha}\ \emph {et~al.}(2022)\citenamefont {Sinha},
  \citenamefont {Murakami},\ and\ \citenamefont {Basu}}]{p_sinha}%
  \BibitemOpen
  \bibfield  {author} {\bibinfo {author} {\bibfnamefont {P.}~\bibnamefont
  {Sinha}}, \bibinfo {author} {\bibfnamefont {S.}~\bibnamefont {Murakami}},\
  and\ \bibinfo {author} {\bibfnamefont {S.}~\bibnamefont {Basu}},\ }\bibfield
  {title} {\bibinfo {title} {Landau levels and magneto-optical transport
  properties of a semi-dirac system},\ }\href
  {https://doi.org/10.1103/PhysRevB.105.205407} {\bibfield  {journal} {\bibinfo
   {journal} {Phys. Rev. B}\ }\textbf {\bibinfo {volume} {105}},\ \bibinfo
  {pages} {205407} (\bibinfo {year} {2022})}\BibitemShut {NoStop}%
\bibitem [{\citenamefont {Mawrie}\ and\ \citenamefont
  {Muralidharan}(2019)}]{Alestinmawrie}%
  \BibitemOpen
  \bibfield  {author} {\bibinfo {author} {\bibfnamefont {A.}~\bibnamefont
  {Mawrie}}\ and\ \bibinfo {author} {\bibfnamefont {B.}~\bibnamefont
  {Muralidharan}},\ }\bibfield  {title} {\bibinfo {title} {Direction-dependent
  giant optical conductivity in two-dimensional semi-dirac materials},\ }\href
  {https://doi.org/10.1103/PhysRevB.99.075415} {\bibfield  {journal} {\bibinfo
  {journal} {Phys. Rev. B}\ }\textbf {\bibinfo {volume} {99}},\ \bibinfo
  {pages} {075415} (\bibinfo {year} {2019})}\BibitemShut {NoStop}%
\bibitem [{\citenamefont {Carbotte}\ and\ \citenamefont
  {Nicol}(2019)}]{carbotte2019}%
  \BibitemOpen
  \bibfield  {author} {\bibinfo {author} {\bibfnamefont {J.~P.}\ \bibnamefont
  {Carbotte}}\ and\ \bibinfo {author} {\bibfnamefont {E.~J.}\ \bibnamefont
  {Nicol}},\ }\bibfield  {title} {\bibinfo {title} {Signatures of merging dirac
  points in optics and transport},\ }\href
  {https://doi.org/10.1103/PhysRevB.100.035441} {\bibfield  {journal} {\bibinfo
   {journal} {Phys. Rev. B}\ }\textbf {\bibinfo {volume} {100}},\ \bibinfo
  {pages} {035441} (\bibinfo {year} {2019})}\BibitemShut {NoStop}%
\bibitem [{\citenamefont {Oriekhov}\ and\ \citenamefont
  {Gusynin}(2022)}]{OriekhovGusynin}%
  \BibitemOpen
  \bibfield  {author} {\bibinfo {author} {\bibfnamefont {D.~O.}\ \bibnamefont
  {Oriekhov}}\ and\ \bibinfo {author} {\bibfnamefont {V.~P.}\ \bibnamefont
  {Gusynin}},\ }\bibfield  {title} {\bibinfo {title} {Optical conductivity of
  semi-dirac and pseudospin-1 models: Zitterbewegung approach},\ }\href
  {https://doi.org/10.1103/PhysRevB.106.115143} {\bibfield  {journal} {\bibinfo
   {journal} {Phys. Rev. B}\ }\textbf {\bibinfo {volume} {106}},\ \bibinfo
  {pages} {115143} (\bibinfo {year} {2022})}\BibitemShut {NoStop}%
\bibitem [{\citenamefont {Xiong}\ \emph {et~al.}(2023)\citenamefont {Xiong},
  \citenamefont {Ba}, \citenamefont {Duan}, \citenamefont {Deng}, \citenamefont
  {Wang},\ and\ \citenamefont {Wang}}]{Xiong2023}%
  \BibitemOpen
  \bibfield  {author} {\bibinfo {author} {\bibfnamefont {Q.-Y.}\ \bibnamefont
  {Xiong}}, \bibinfo {author} {\bibfnamefont {J.-Y.}\ \bibnamefont {Ba}},
  \bibinfo {author} {\bibfnamefont {H.-J.}\ \bibnamefont {Duan}}, \bibinfo
  {author} {\bibfnamefont {M.-X.}\ \bibnamefont {Deng}}, \bibinfo {author}
  {\bibfnamefont {Y.-M.}\ \bibnamefont {Wang}},\ and\ \bibinfo {author}
  {\bibfnamefont {R.-Q.}\ \bibnamefont {Wang}},\ }\bibfield  {title} {\bibinfo
  {title} {Optical conductivity and polarization rotation of type-ii semi-dirac
  materials},\ }\href {https://doi.org/10.1103/PhysRevB.107.155150} {\bibfield
  {journal} {\bibinfo  {journal} {Phys. Rev. B}\ }\textbf {\bibinfo {volume}
  {107}},\ \bibinfo {pages} {155150} (\bibinfo {year} {2023})}\BibitemShut
  {NoStop}%
\bibitem [{\citenamefont {Ghosh}\ \emph {et~al.}(2025)\citenamefont {Ghosh},
  \citenamefont {Bandyopadhyay},\ and\ \citenamefont {Singh}}]{Ashutosh25}%
  \BibitemOpen
  \bibfield  {author} {\bibinfo {author} {\bibfnamefont {B.}~\bibnamefont
  {Ghosh}}, \bibinfo {author} {\bibfnamefont {M.}~\bibnamefont
  {Bandyopadhyay}},\ and\ \bibinfo {author} {\bibfnamefont {A.}~\bibnamefont
  {Singh}},\ }\bibfield  {title} {\bibinfo {title} {Optical pumping controls
  anisotropic response in semi-dirac systems},\ }\href
  {https://doi.org/10.1103/dfbv-9vcj} {\bibfield  {journal} {\bibinfo
  {journal} {Phys. Rev. B}\ }\textbf {\bibinfo {volume} {111}},\ \bibinfo
  {pages} {245414} (\bibinfo {year} {2025})}\BibitemShut {NoStop}%
\bibitem [{\citenamefont {Narayan}(2015)}]{awdesh}%
  \BibitemOpen
  \bibfield  {author} {\bibinfo {author} {\bibfnamefont {A.}~\bibnamefont
  {Narayan}},\ }\bibfield  {title} {\bibinfo {title} {Floquet dynamics in
  two-dimensional semi-dirac semimetals and three-dimensional dirac
  semimetals},\ }\href {https://doi.org/10.1103/PhysRevB.91.205445} {\bibfield
  {journal} {\bibinfo  {journal} {Phys. Rev. B}\ }\textbf {\bibinfo {volume}
  {91}},\ \bibinfo {pages} {205445} (\bibinfo {year} {2015})}\BibitemShut
  {NoStop}%
\bibitem [{\citenamefont {Islam}\ and\ \citenamefont
  {Saha}(2018)}]{SKF_Arijit}%
  \BibitemOpen
  \bibfield  {author} {\bibinfo {author} {\bibfnamefont {S.~F.}\ \bibnamefont
  {Islam}}\ and\ \bibinfo {author} {\bibfnamefont {A.}~\bibnamefont {Saha}},\
  }\bibfield  {title} {\bibinfo {title} {Driven conductance of an irradiated
  semi-dirac material},\ }\href {https://doi.org/10.1103/PhysRevB.98.235424}
  {\bibfield  {journal} {\bibinfo  {journal} {Phys. Rev. B}\ }\textbf {\bibinfo
  {volume} {98}},\ \bibinfo {pages} {235424} (\bibinfo {year}
  {2018})}\BibitemShut {NoStop}%
\bibitem [{\citenamefont {Chen}\ \emph {et~al.}(2022)\citenamefont {Chen},
  \citenamefont {Yang}, \citenamefont {Zhou}, \citenamefont {Wu}, \citenamefont
  {Duan}, \citenamefont {Deng},\ and\ \citenamefont {Wang}}]{Chen}%
  \BibitemOpen
  \bibfield  {author} {\bibinfo {author} {\bibfnamefont {J.-N.}\ \bibnamefont
  {Chen}}, \bibinfo {author} {\bibfnamefont {Y.-Y.}\ \bibnamefont {Yang}},
  \bibinfo {author} {\bibfnamefont {Y.-L.}\ \bibnamefont {Zhou}}, \bibinfo
  {author} {\bibfnamefont {Y.-J.}\ \bibnamefont {Wu}}, \bibinfo {author}
  {\bibfnamefont {H.-J.}\ \bibnamefont {Duan}}, \bibinfo {author}
  {\bibfnamefont {M.-X.}\ \bibnamefont {Deng}},\ and\ \bibinfo {author}
  {\bibfnamefont {R.-Q.}\ \bibnamefont {Wang}},\ }\bibfield  {title} {\bibinfo
  {title} {Photon-modulated linear and nonlinear anomalous hall effects in
  type-ii semi-dirac semimetals},\ }\href
  {https://doi.org/10.1103/PhysRevB.105.085124} {\bibfield  {journal} {\bibinfo
   {journal} {Phys. Rev. B}\ }\textbf {\bibinfo {volume} {105}},\ \bibinfo
  {pages} {085124} (\bibinfo {year} {2022})}\BibitemShut {NoStop}%
\bibitem [{\citenamefont {Islam}(2024)}]{SKF_valkov}%
  \BibitemOpen
  \bibfield  {author} {\bibinfo {author} {\bibfnamefont {S.~F.}\ \bibnamefont
  {Islam}},\ }\bibfield  {title} {\bibinfo {title} {Photoinduced metallic
  volkov-pankratov states in a semi-dirac material},\ }\href
  {https://doi.org/10.1103/PhysRevB.109.235416} {\bibfield  {journal} {\bibinfo
   {journal} {Phys. Rev. B}\ }\textbf {\bibinfo {volume} {109}},\ \bibinfo
  {pages} {235416} (\bibinfo {year} {2024})}\BibitemShut {NoStop}%
\bibitem [{\citenamefont {Li}\ \emph {et~al.}(2022)\citenamefont {Li},
  \citenamefont {Hu},\ and\ \citenamefont {Ouyang}}]{Li_2022}%
  \BibitemOpen
  \bibfield  {author} {\bibinfo {author} {\bibfnamefont {H.}~\bibnamefont
  {Li}}, \bibinfo {author} {\bibfnamefont {X.}~\bibnamefont {Hu}},\ and\
  \bibinfo {author} {\bibfnamefont {G.}~\bibnamefont {Ouyang}},\ }\bibfield
  {title} {\bibinfo {title} {Orientation-dependent crossover from retro to
  specular andreev reflections in semi-dirac materials},\ }\href
  {https://doi.org/10.1088/1367-2630/ac6e7f} {\bibfield  {journal} {\bibinfo
  {journal} {New J. Phys.}\ }\textbf {\bibinfo {volume} {24}},\ \bibinfo
  {pages} {053049} (\bibinfo {year} {2022})}\BibitemShut {NoStop}%
\bibitem [{\citenamefont {Ross-Harvey}\ \emph {et~al.}(2025)\citenamefont
  {Ross-Harvey}, \citenamefont {Iurov}, \citenamefont {Zhemchuzhna},
  \citenamefont {Gumbs}, \citenamefont {Huang},\ and\ \citenamefont
  {Fekete}}]{A_lurov}%
  \BibitemOpen
  \bibfield  {author} {\bibinfo {author} {\bibfnamefont {G.}~\bibnamefont
  {Ross-Harvey}}, \bibinfo {author} {\bibfnamefont {A.}~\bibnamefont {Iurov}},
  \bibinfo {author} {\bibfnamefont {L.}~\bibnamefont {Zhemchuzhna}}, \bibinfo
  {author} {\bibfnamefont {G.}~\bibnamefont {Gumbs}}, \bibinfo {author}
  {\bibfnamefont {D.}~\bibnamefont {Huang}},\ and\ \bibinfo {author}
  {\bibfnamefont {P.}~\bibnamefont {Fekete}},\ }\bibfield  {title} {\bibinfo
  {title} {Dynamical polarization function, anisotropic plasmon modes, and
  dephasing rates for interacting electrons in semi-dirac bands},\ }\href
  {https://doi.org/10.1103/PhysRevB.111.045413} {\bibfield  {journal} {\bibinfo
   {journal} {Phys. Rev. B}\ }\textbf {\bibinfo {volume} {111}},\ \bibinfo
  {pages} {045413} (\bibinfo {year} {2025})}\BibitemShut {NoStop}%
\bibitem [{\citenamefont {Nezafat}\ \emph {et~al.}(2025)\citenamefont
  {Nezafat}, \citenamefont {Barati},\ and\ \citenamefont
  {Abedinpour}}]{abedinpour}%
  \BibitemOpen
  \bibfield  {author} {\bibinfo {author} {\bibfnamefont {M.}~\bibnamefont
  {Nezafat}}, \bibinfo {author} {\bibfnamefont {S.}~\bibnamefont {Barati}},\
  and\ \bibinfo {author} {\bibfnamefont {S.~H.}\ \bibnamefont {Abedinpour}},\
  }\bibfield  {title} {\bibinfo {title} {Magnetothermoelectricity of
  anisotropic two-dimensional materials},\ }\href
  {https://doi.org/10.1103/PhysRevB.111.165423} {\bibfield  {journal} {\bibinfo
   {journal} {Phys. Rev. B}\ }\textbf {\bibinfo {volume} {111}},\ \bibinfo
  {pages} {165423} (\bibinfo {year} {2025})}\BibitemShut {NoStop}%
\bibitem [{\citenamefont {Elsayed}\ \emph {et~al.}(2025)\citenamefont
  {Elsayed}, \citenamefont {Uchoa},\ and\ \citenamefont {Kotov}}]{elsayed}%
  \BibitemOpen
  \bibfield  {author} {\bibinfo {author} {\bibfnamefont {M.~M.}\ \bibnamefont
  {Elsayed}}, \bibinfo {author} {\bibfnamefont {B.}~\bibnamefont {Uchoa}},\
  and\ \bibinfo {author} {\bibfnamefont {V.~N.}\ \bibnamefont {Kotov}},\
  }\bibfield  {title} {\bibinfo {title} {Coulomb interactions in systems of
  generalized semi-dirac fermions},\ }\href
  {https://doi.org/10.1103/PhysRevB.111.165127} {\bibfield  {journal} {\bibinfo
   {journal} {Phys. Rev. B}\ }\textbf {\bibinfo {volume} {111}},\ \bibinfo
  {pages} {165127} (\bibinfo {year} {2025})}\BibitemShut {NoStop}%
\bibitem [{\citenamefont {Zabolotskiy}\ and\ \citenamefont
  {Lozovik}(2016)}]{lozovik}%
  \BibitemOpen
  \bibfield  {author} {\bibinfo {author} {\bibfnamefont {A.~D.}\ \bibnamefont
  {Zabolotskiy}}\ and\ \bibinfo {author} {\bibfnamefont {Y.~E.}\ \bibnamefont
  {Lozovik}},\ }\bibfield  {title} {\bibinfo {title} {Strain-induced
  pseudomagnetic field in the dirac semimetal borophene},\ }\href
  {https://doi.org/10.1103/PhysRevB.94.165403} {\bibfield  {journal} {\bibinfo
  {journal} {Phys. Rev. B}\ }\textbf {\bibinfo {volume} {94}},\ \bibinfo
  {pages} {165403} (\bibinfo {year} {2016})}\BibitemShut {NoStop}%
\bibitem [{\citenamefont {Lv}\ \emph {et~al.}(2015)\citenamefont {Lv},
  \citenamefont {Weng}, \citenamefont {Fu}, \citenamefont {Wang}, \citenamefont
  {Miao}, \citenamefont {Ma}, \citenamefont {Richard}, \citenamefont {Huang},
  \citenamefont {Zhao}, \citenamefont {Chen}, \citenamefont {Fang},
  \citenamefont {Dai}, \citenamefont {Qian},\ and\ \citenamefont
  {Ding}}]{weylTaAS_2015}%
  \BibitemOpen
  \bibfield  {author} {\bibinfo {author} {\bibfnamefont {B.~Q.}\ \bibnamefont
  {Lv}}, \bibinfo {author} {\bibfnamefont {H.~M.}\ \bibnamefont {Weng}},
  \bibinfo {author} {\bibfnamefont {B.~B.}\ \bibnamefont {Fu}}, \bibinfo
  {author} {\bibfnamefont {X.~P.}\ \bibnamefont {Wang}}, \bibinfo {author}
  {\bibfnamefont {H.}~\bibnamefont {Miao}}, \bibinfo {author} {\bibfnamefont
  {J.}~\bibnamefont {Ma}}, \bibinfo {author} {\bibfnamefont {P.}~\bibnamefont
  {Richard}}, \bibinfo {author} {\bibfnamefont {X.~C.}\ \bibnamefont {Huang}},
  \bibinfo {author} {\bibfnamefont {L.~X.}\ \bibnamefont {Zhao}}, \bibinfo
  {author} {\bibfnamefont {G.~F.}\ \bibnamefont {Chen}}, \bibinfo {author}
  {\bibfnamefont {Z.}~\bibnamefont {Fang}}, \bibinfo {author} {\bibfnamefont
  {X.}~\bibnamefont {Dai}}, \bibinfo {author} {\bibfnamefont {T.}~\bibnamefont
  {Qian}},\ and\ \bibinfo {author} {\bibfnamefont {H.}~\bibnamefont {Ding}},\
  }\bibfield  {title} {\bibinfo {title} {Experimental discovery of weyl
  semimetal taas},\ }\href {https://doi.org/10.1103/PhysRevX.5.031013}
  {\bibfield  {journal} {\bibinfo  {journal} {Phys. Rev. X}\ }\textbf {\bibinfo
  {volume} {5}},\ \bibinfo {pages} {031013} (\bibinfo {year}
  {2015})}\BibitemShut {NoStop}%
\bibitem [{\citenamefont {Deng}\ \emph {et~al.}(2016)\citenamefont {Deng},
  \citenamefont {Wan}, \citenamefont {Deng}, \citenamefont {Zhang},
  \citenamefont {Ding}, \citenamefont {Wang}, \citenamefont {Yan},
  \citenamefont {Huang}, \citenamefont {Zhang}, \citenamefont {Xu},
  \citenamefont {Denlinger}, \citenamefont {Fedorov}, \citenamefont {Yang},
  \citenamefont {Duan}, \citenamefont {Yao}, \citenamefont {Wu}, \citenamefont
  {Fan}, \citenamefont {Zhang}, \citenamefont {Chen},\ and\ \citenamefont
  {Zhou}}]{Deng_2016}%
  \BibitemOpen
  \bibfield  {author} {\bibinfo {author} {\bibfnamefont {K.}~\bibnamefont
  {Deng}}, \bibinfo {author} {\bibfnamefont {G.}~\bibnamefont {Wan}}, \bibinfo
  {author} {\bibfnamefont {P.}~\bibnamefont {Deng}}, \bibinfo {author}
  {\bibfnamefont {K.}~\bibnamefont {Zhang}}, \bibinfo {author} {\bibfnamefont
  {S.}~\bibnamefont {Ding}}, \bibinfo {author} {\bibfnamefont {E.}~\bibnamefont
  {Wang}}, \bibinfo {author} {\bibfnamefont {M.}~\bibnamefont {Yan}}, \bibinfo
  {author} {\bibfnamefont {H.}~\bibnamefont {Huang}}, \bibinfo {author}
  {\bibfnamefont {H.}~\bibnamefont {Zhang}}, \bibinfo {author} {\bibfnamefont
  {Z.}~\bibnamefont {Xu}}, \bibinfo {author} {\bibfnamefont {J.}~\bibnamefont
  {Denlinger}}, \bibinfo {author} {\bibfnamefont {A.}~\bibnamefont {Fedorov}},
  \bibinfo {author} {\bibfnamefont {H.}~\bibnamefont {Yang}}, \bibinfo {author}
  {\bibfnamefont {W.}~\bibnamefont {Duan}}, \bibinfo {author} {\bibfnamefont
  {H.}~\bibnamefont {Yao}}, \bibinfo {author} {\bibfnamefont {Y.}~\bibnamefont
  {Wu}}, \bibinfo {author} {\bibfnamefont {S.}~\bibnamefont {Fan}}, \bibinfo
  {author} {\bibfnamefont {H.}~\bibnamefont {Zhang}}, \bibinfo {author}
  {\bibfnamefont {X.}~\bibnamefont {Chen}},\ and\ \bibinfo {author}
  {\bibfnamefont {S.}~\bibnamefont {Zhou}},\ }\bibfield  {title} {\bibinfo
  {title} {Experimental observation of topological fermi arcs in type-ii weyl
  semimetal mote$_2$},\ }\href {https://doi.org/10.1038/nphys3871} {\bibfield
  {journal} {\bibinfo  {journal} {Nature Physics}\ }\textbf {\bibinfo {volume}
  {12}},\ \bibinfo {pages} {1105} (\bibinfo {year} {2016})}\BibitemShut
  {NoStop}%
\bibitem [{\citenamefont {Wu}\ \emph {et~al.}(2016)\citenamefont {Wu},
  \citenamefont {Mou}, \citenamefont {Jo}, \citenamefont {Sun}, \citenamefont
  {Huang}, \citenamefont {Bud'ko}, \citenamefont {Canfield},\ and\
  \citenamefont {Kaminski}}]{weyltype2_2016}%
  \BibitemOpen
  \bibfield  {author} {\bibinfo {author} {\bibfnamefont {Y.}~\bibnamefont
  {Wu}}, \bibinfo {author} {\bibfnamefont {D.}~\bibnamefont {Mou}}, \bibinfo
  {author} {\bibfnamefont {N.~H.}\ \bibnamefont {Jo}}, \bibinfo {author}
  {\bibfnamefont {K.}~\bibnamefont {Sun}}, \bibinfo {author} {\bibfnamefont
  {L.}~\bibnamefont {Huang}}, \bibinfo {author} {\bibfnamefont {S.~L.}\
  \bibnamefont {Bud'ko}}, \bibinfo {author} {\bibfnamefont {P.~C.}\
  \bibnamefont {Canfield}},\ and\ \bibinfo {author} {\bibfnamefont
  {A.}~\bibnamefont {Kaminski}},\ }\bibfield  {title} {\bibinfo {title}
  {Observation of fermi arcs in the type-ii weyl semimetal candidate
  ${\mathrm{wte}}_{2}$},\ }\href {https://doi.org/10.1103/PhysRevB.94.121113}
  {\bibfield  {journal} {\bibinfo  {journal} {Phys. Rev. B}\ }\textbf {\bibinfo
  {volume} {94}},\ \bibinfo {pages} {121113} (\bibinfo {year}
  {2016})}\BibitemShut {NoStop}%
\bibitem [{\citenamefont {Sadhukhan}\ and\ \citenamefont
  {Agarwal}(2017)}]{amit2017}%
  \BibitemOpen
  \bibfield  {author} {\bibinfo {author} {\bibfnamefont {K.}~\bibnamefont
  {Sadhukhan}}\ and\ \bibinfo {author} {\bibfnamefont {A.}~\bibnamefont
  {Agarwal}},\ }\bibfield  {title} {\bibinfo {title} {Anisotropic plasmons,
  friedel oscillations, and screening in $8\text{\ensuremath{-}}pmmn$
  borophene},\ }\href {https://doi.org/10.1103/PhysRevB.96.035410} {\bibfield
  {journal} {\bibinfo  {journal} {Phys. Rev. B}\ }\textbf {\bibinfo {volume}
  {96}},\ \bibinfo {pages} {035410} (\bibinfo {year} {2017})}\BibitemShut
  {NoStop}%
\bibitem [{\citenamefont {Torbatian}\ \emph {et~al.}(2021)\citenamefont
  {Torbatian}, \citenamefont {Novko},\ and\ \citenamefont
  {Asgari}}]{reza_2021}%
  \BibitemOpen
  \bibfield  {author} {\bibinfo {author} {\bibfnamefont {Z.}~\bibnamefont
  {Torbatian}}, \bibinfo {author} {\bibfnamefont {D.}~\bibnamefont {Novko}},\
  and\ \bibinfo {author} {\bibfnamefont {R.}~\bibnamefont {Asgari}},\
  }\bibfield  {title} {\bibinfo {title} {Hyperbolic plasmon modes in tilted
  dirac cone phases of borophene},\ }\href
  {https://doi.org/10.1103/PhysRevB.104.075432} {\bibfield  {journal} {\bibinfo
   {journal} {Phys. Rev. B}\ }\textbf {\bibinfo {volume} {104}},\ \bibinfo
  {pages} {075432} (\bibinfo {year} {2021})}\BibitemShut {NoStop}%
\bibitem [{\citenamefont {Verma}\ \emph {et~al.}(2017)\citenamefont {Verma},
  \citenamefont {Mawrie},\ and\ \citenamefont {Ghosh}}]{alestin_borophene}%
  \BibitemOpen
  \bibfield  {author} {\bibinfo {author} {\bibfnamefont {S.}~\bibnamefont
  {Verma}}, \bibinfo {author} {\bibfnamefont {A.}~\bibnamefont {Mawrie}},\ and\
  \bibinfo {author} {\bibfnamefont {T.~K.}\ \bibnamefont {Ghosh}},\ }\bibfield
  {title} {\bibinfo {title} {Effect of electron-hole asymmetry on optical
  conductivity in $8\ensuremath{-}pmmn$ borophene},\ }\href
  {https://doi.org/10.1103/PhysRevB.96.155418} {\bibfield  {journal} {\bibinfo
  {journal} {Phys. Rev. B}\ }\textbf {\bibinfo {volume} {96}},\ \bibinfo
  {pages} {155418} (\bibinfo {year} {2017})}\BibitemShut {NoStop}%
\bibitem [{\citenamefont {Mojarro}\ \emph {et~al.}(2021)\citenamefont
  {Mojarro}, \citenamefont {Carrillo-Bastos},\ and\ \citenamefont
  {Maytorena}}]{mojarro}%
  \BibitemOpen
  \bibfield  {author} {\bibinfo {author} {\bibfnamefont {M.~A.}\ \bibnamefont
  {Mojarro}}, \bibinfo {author} {\bibfnamefont {R.}~\bibnamefont
  {Carrillo-Bastos}},\ and\ \bibinfo {author} {\bibfnamefont {J.~A.}\
  \bibnamefont {Maytorena}},\ }\bibfield  {title} {\bibinfo {title} {Optical
  properties of massive anisotropic tilted dirac systems},\ }\href
  {https://doi.org/10.1103/PhysRevB.103.165415} {\bibfield  {journal} {\bibinfo
   {journal} {Phys. Rev. B}\ }\textbf {\bibinfo {volume} {103}},\ \bibinfo
  {pages} {165415} (\bibinfo {year} {2021})}\BibitemShut {NoStop}%
\bibitem [{\citenamefont {Islam}\ and\ \citenamefont
  {Jayannavar}(2017)}]{SKF_jayannavar}%
  \BibitemOpen
  \bibfield  {author} {\bibinfo {author} {\bibfnamefont {S.~F.}\ \bibnamefont
  {Islam}}\ and\ \bibinfo {author} {\bibfnamefont {A.~M.}\ \bibnamefont
  {Jayannavar}},\ }\bibfield  {title} {\bibinfo {title} {Signature of tilted
  dirac cones in weiss oscillations of $8\ensuremath{-}pmmn$ borophene},\
  }\href {https://doi.org/10.1103/PhysRevB.96.235405} {\bibfield  {journal}
  {\bibinfo  {journal} {Phys. Rev. B}\ }\textbf {\bibinfo {volume} {96}},\
  \bibinfo {pages} {235405} (\bibinfo {year} {2017})}\BibitemShut {NoStop}%
\bibitem [{\citenamefont {Islam}(2018)}]{Islam_2018_jpcm}%
  \BibitemOpen
  \bibfield  {author} {\bibinfo {author} {\bibfnamefont {S.~F.}\ \bibnamefont
  {Islam}},\ }\bibfield  {title} {\bibinfo {title} {Magnetotransport properties
  of 8-pmmn borophene: effects of hall field and strain},\ }\href
  {https://doi.org/10.1088/1361-648X/aac8b3} {\bibfield  {journal} {\bibinfo
  {journal} {J.Phys. : Condens. Matter}\ }\textbf {\bibinfo {volume} {30}},\
  \bibinfo {pages} {275301} (\bibinfo {year} {2018})}\BibitemShut {NoStop}%
\bibitem [{\citenamefont {Xu}\ \emph {et~al.}(2016)\citenamefont {Xu},
  \citenamefont {Dai}, \citenamefont {Zhao}, \citenamefont {Wang},
  \citenamefont {Yang}, \citenamefont {Zhang}, \citenamefont {Liu},
  \citenamefont {Xiao}, \citenamefont {Chen}, \citenamefont {Taylor},
  \citenamefont {Yarotski}, \citenamefont {Prasankumar},\ and\ \citenamefont
  {Qiu}}]{xu_optic}%
  \BibitemOpen
  \bibfield  {author} {\bibinfo {author} {\bibfnamefont {B.}~\bibnamefont
  {Xu}}, \bibinfo {author} {\bibfnamefont {Y.~M.}\ \bibnamefont {Dai}},
  \bibinfo {author} {\bibfnamefont {L.~X.}\ \bibnamefont {Zhao}}, \bibinfo
  {author} {\bibfnamefont {K.}~\bibnamefont {Wang}}, \bibinfo {author}
  {\bibfnamefont {R.}~\bibnamefont {Yang}}, \bibinfo {author} {\bibfnamefont
  {W.}~\bibnamefont {Zhang}}, \bibinfo {author} {\bibfnamefont {J.~Y.}\
  \bibnamefont {Liu}}, \bibinfo {author} {\bibfnamefont {H.}~\bibnamefont
  {Xiao}}, \bibinfo {author} {\bibfnamefont {G.~F.}\ \bibnamefont {Chen}},
  \bibinfo {author} {\bibfnamefont {A.~J.}\ \bibnamefont {Taylor}}, \bibinfo
  {author} {\bibfnamefont {D.~A.}\ \bibnamefont {Yarotski}}, \bibinfo {author}
  {\bibfnamefont {R.~P.}\ \bibnamefont {Prasankumar}},\ and\ \bibinfo {author}
  {\bibfnamefont {X.~G.}\ \bibnamefont {Qiu}},\ }\bibfield  {title} {\bibinfo
  {title} {Optical spectroscopy of the weyl semimetal taas},\ }\href
  {https://doi.org/10.1103/PhysRevB.93.121110} {\bibfield  {journal} {\bibinfo
  {journal} {Phys. Rev. B}\ }\textbf {\bibinfo {volume} {93}},\ \bibinfo
  {pages} {121110} (\bibinfo {year} {2016})}\BibitemShut {NoStop}%
\bibitem [{\citenamefont {Chen}\ \emph {et~al.}(2019)\citenamefont {Chen},
  \citenamefont {Kutayiah}, \citenamefont {Oladyshkin}, \citenamefont
  {Tokman},\ and\ \citenamefont {Belyanin}}]{Q_chen}%
  \BibitemOpen
  \bibfield  {author} {\bibinfo {author} {\bibfnamefont {Q.}~\bibnamefont
  {Chen}}, \bibinfo {author} {\bibfnamefont {A.~R.}\ \bibnamefont {Kutayiah}},
  \bibinfo {author} {\bibfnamefont {I.}~\bibnamefont {Oladyshkin}}, \bibinfo
  {author} {\bibfnamefont {M.}~\bibnamefont {Tokman}},\ and\ \bibinfo {author}
  {\bibfnamefont {A.}~\bibnamefont {Belyanin}},\ }\bibfield  {title} {\bibinfo
  {title} {Optical properties and electromagnetic modes of weyl semimetals},\
  }\href {https://doi.org/10.1103/PhysRevB.99.075137} {\bibfield  {journal}
  {\bibinfo  {journal} {Phys. Rev. B}\ }\textbf {\bibinfo {volume} {99}},\
  \bibinfo {pages} {075137} (\bibinfo {year} {2019})}\BibitemShut {NoStop}%
\bibitem [{\citenamefont {Zyuzin}(2017)}]{Vladmir_Zyuzin}%
  \BibitemOpen
  \bibfield  {author} {\bibinfo {author} {\bibfnamefont {V.~A.}\ \bibnamefont
  {Zyuzin}},\ }\bibfield  {title} {\bibinfo {title} {Magnetotransport of weyl
  semimetals due to the chiral anomaly},\ }\href
  {https://doi.org/10.1103/PhysRevB.95.245128} {\bibfield  {journal} {\bibinfo
  {journal} {Phys. Rev. B}\ }\textbf {\bibinfo {volume} {95}},\ \bibinfo
  {pages} {245128} (\bibinfo {year} {2017})}\BibitemShut {NoStop}%
\bibitem [{\citenamefont {Sharma}\ \emph {et~al.}(2017)\citenamefont {Sharma},
  \citenamefont {Goswami},\ and\ \citenamefont {Tewari}}]{Sharma_goswami}%
  \BibitemOpen
  \bibfield  {author} {\bibinfo {author} {\bibfnamefont {G.}~\bibnamefont
  {Sharma}}, \bibinfo {author} {\bibfnamefont {P.}~\bibnamefont {Goswami}},\
  and\ \bibinfo {author} {\bibfnamefont {S.}~\bibnamefont {Tewari}},\
  }\bibfield  {title} {\bibinfo {title} {Chiral anomaly and longitudinal
  magnetotransport in type-ii weyl semimetals},\ }\href
  {https://doi.org/10.1103/PhysRevB.96.045112} {\bibfield  {journal} {\bibinfo
  {journal} {Phys. Rev. B}\ }\textbf {\bibinfo {volume} {96}},\ \bibinfo
  {pages} {045112} (\bibinfo {year} {2017})}\BibitemShut {NoStop}%
\bibitem [{\citenamefont {Wei}\ \emph {et~al.}(2018)\citenamefont {Wei},
  \citenamefont {Li}, \citenamefont {Qi},\ and\ \citenamefont {Feng}}]{Wei_yi}%
  \BibitemOpen
  \bibfield  {author} {\bibinfo {author} {\bibfnamefont {Y.-W.}\ \bibnamefont
  {Wei}}, \bibinfo {author} {\bibfnamefont {C.-K.}\ \bibnamefont {Li}},
  \bibinfo {author} {\bibfnamefont {J.}~\bibnamefont {Qi}},\ and\ \bibinfo
  {author} {\bibfnamefont {J.}~\bibnamefont {Feng}},\ }\bibfield  {title}
  {\bibinfo {title} {Magnetoconductivity of type-ii weyl semimetals},\ }\href
  {https://doi.org/10.1103/PhysRevB.97.205131} {\bibfield  {journal} {\bibinfo
  {journal} {Phys. Rev. B}\ }\textbf {\bibinfo {volume} {97}},\ \bibinfo
  {pages} {205131} (\bibinfo {year} {2018})}\BibitemShut {NoStop}%
\bibitem [{\citenamefont {Das}\ and\ \citenamefont
  {Agarwal}(2019)}]{Das_kamal}%
  \BibitemOpen
  \bibfield  {author} {\bibinfo {author} {\bibfnamefont {K.}~\bibnamefont
  {Das}}\ and\ \bibinfo {author} {\bibfnamefont {A.}~\bibnamefont {Agarwal}},\
  }\bibfield  {title} {\bibinfo {title} {Linear magnetochiral transport in
  tilted type-i and type-ii weyl semimetals},\ }\href
  {https://doi.org/10.1103/PhysRevB.99.085405} {\bibfield  {journal} {\bibinfo
  {journal} {Phys. Rev. B}\ }\textbf {\bibinfo {volume} {99}},\ \bibinfo
  {pages} {085405} (\bibinfo {year} {2019})}\BibitemShut {NoStop}%
\bibitem [{\citenamefont {Yan}\ \emph {et~al.}(2023)\citenamefont {Yan},
  \citenamefont {Tan}, \citenamefont {Guo},\ and\ \citenamefont
  {Chang}}]{xu_yan2023}%
  \BibitemOpen
  \bibfield  {author} {\bibinfo {author} {\bibfnamefont {C.-X.}\ \bibnamefont
  {Yan}}, \bibinfo {author} {\bibfnamefont {C.-Y.}\ \bibnamefont {Tan}},
  \bibinfo {author} {\bibfnamefont {H.}~\bibnamefont {Guo}},\ and\ \bibinfo
  {author} {\bibfnamefont {H.-R.}\ \bibnamefont {Chang}},\ }\bibfield  {title}
  {\bibinfo {title} {Highly anisotropic optical conductivities in
  two-dimensional tilted semi-dirac bands},\ }\href
  {https://doi.org/10.1103/PhysRevB.108.195427} {\bibfield  {journal} {\bibinfo
   {journal} {Phys. Rev. B}\ }\textbf {\bibinfo {volume} {108}},\ \bibinfo
  {pages} {195427} (\bibinfo {year} {2023})}\BibitemShut {NoStop}%
\bibitem [{\citenamefont {Zyuzin}\ \emph {et~al.}(2012)\citenamefont {Zyuzin},
  \citenamefont {Wu},\ and\ \citenamefont {Burkov}}]{zyuzin}%
  \BibitemOpen
  \bibfield  {author} {\bibinfo {author} {\bibfnamefont {A.~A.}\ \bibnamefont
  {Zyuzin}}, \bibinfo {author} {\bibfnamefont {S.}~\bibnamefont {Wu}},\ and\
  \bibinfo {author} {\bibfnamefont {A.~A.}\ \bibnamefont {Burkov}},\ }\bibfield
   {title} {\bibinfo {title} {Weyl semimetal with broken time reversal and
  inversion symmetries},\ }\href {https://doi.org/10.1103/PhysRevB.85.165110}
  {\bibfield  {journal} {\bibinfo  {journal} {Phys. Rev. B}\ }\textbf {\bibinfo
  {volume} {85}},\ \bibinfo {pages} {165110} (\bibinfo {year}
  {2012})}\BibitemShut {NoStop}%
\bibitem [{\citenamefont {Zyuzin}\ and\ \citenamefont
  {Zyuzin}(2019)}]{PhysRevB.100.121402}%
  \BibitemOpen
  \bibfield  {author} {\bibinfo {author} {\bibfnamefont {A.~A.}\ \bibnamefont
  {Zyuzin}}\ and\ \bibinfo {author} {\bibfnamefont {A.~Y.}\ \bibnamefont
  {Zyuzin}},\ }\bibfield  {title} {\bibinfo {title} {Spin and valley waves in
  dirac semimetals with population imbalance},\ }\href
  {https://doi.org/10.1103/PhysRevB.100.121402} {\bibfield  {journal} {\bibinfo
   {journal} {Phys. Rev. B}\ }\textbf {\bibinfo {volume} {100}},\ \bibinfo
  {pages} {121402} (\bibinfo {year} {2019})}\BibitemShut {NoStop}%
\bibitem [{\citenamefont {Oka}\ and\ \citenamefont
  {Aoki}(2009)}]{oka2009photovoltaic}%
  \BibitemOpen
  \bibfield  {author} {\bibinfo {author} {\bibfnamefont {T.}~\bibnamefont
  {Oka}}\ and\ \bibinfo {author} {\bibfnamefont {H.}~\bibnamefont {Aoki}},\
  }\bibfield  {title} {\bibinfo {title} {Photovoltaic hall effect in
  graphene},\ }\href@noop {} {\bibfield  {journal} {\bibinfo  {journal} {Phys.
  Rev. B}\ }\textbf {\bibinfo {volume} {79}},\ \bibinfo {pages} {081406}
  (\bibinfo {year} {2009})}\BibitemShut {NoStop}%
\bibitem [{\citenamefont {de~Juan}\ \emph {et~al.}(2017)\citenamefont
  {de~Juan}, \citenamefont {Grushin}, \citenamefont {Morimoto},\ and\
  \citenamefont {Moore}}]{dejuan2017quantized}%
  \BibitemOpen
  \bibfield  {author} {\bibinfo {author} {\bibfnamefont {F.}~\bibnamefont
  {de~Juan}}, \bibinfo {author} {\bibfnamefont {A.~G.}\ \bibnamefont
  {Grushin}}, \bibinfo {author} {\bibfnamefont {T.}~\bibnamefont {Morimoto}},\
  and\ \bibinfo {author} {\bibfnamefont {J.~E.}\ \bibnamefont {Moore}},\
  }\bibfield  {title} {\bibinfo {title} {Quantized circular photogalvanic
  effect in weyl semimetals},\ }\href {https://doi.org/10.1038/ncomms15995}
  {\bibfield  {journal} {\bibinfo  {journal} {Nat. Comm.}\ }\textbf {\bibinfo
  {volume} {8}},\ \bibinfo {pages} {15995} (\bibinfo {year}
  {2017})}\BibitemShut {NoStop}%
\bibitem [{\citenamefont {Bykov}\ \emph {et~al.}(2012)\citenamefont {Bykov},
  \citenamefont {Makarov} \emph {et~al.}}]{bykov2012second}%
  \BibitemOpen
  \bibfield  {author} {\bibinfo {author} {\bibfnamefont {A.~Y.}\ \bibnamefont
  {Bykov}}, \bibinfo {author} {\bibfnamefont {S.~V.}\ \bibnamefont {Makarov}},
  \emph {et~al.},\ }\bibfield  {title} {\bibinfo {title} {Second harmonic
  generation in graphene induced by direct electric current},\ }\href
  {https://doi.org/10.1103/PhysRevB.85.121413} {\bibfield  {journal} {\bibinfo
  {journal} {Phys. Rev. B}\ }\textbf {\bibinfo {volume} {85}},\ \bibinfo
  {pages} {121413} (\bibinfo {year} {2012})}\BibitemShut {NoStop}%
\bibitem [{\citenamefont {Wang}\ \emph {et~al.}(2016)\citenamefont {Wang} \emph
  {et~al.}}]{wang2016giant}%
  \BibitemOpen
  \bibfield  {author} {\bibinfo {author} {\bibfnamefont {H.}~\bibnamefont
  {Wang}} \emph {et~al.},\ }\bibfield  {title} {\bibinfo {title} {Giant
  electric-field-induced second harmonic generation in graphene},\ }\href
  {https://doi.org/10.1021/acs.nanolett.6b00739} {\bibfield  {journal}
  {\bibinfo  {journal} {Nano Letters}\ }\textbf {\bibinfo {volume} {16}},\
  \bibinfo {pages} {3213} (\bibinfo {year} {2016})}\BibitemShut {NoStop}%
\bibitem [{\citenamefont {Xu}\ \emph {et~al.}(2018)\citenamefont {Xu},
  \citenamefont {Ma}, \citenamefont {Basak} \emph {et~al.}}]{Xu2018SHGWeyl}%
  \BibitemOpen
  \bibfield  {author} {\bibinfo {author} {\bibfnamefont {S.-Y.}\ \bibnamefont
  {Xu}}, \bibinfo {author} {\bibfnamefont {Q.}~\bibnamefont {Ma}}, \bibinfo
  {author} {\bibfnamefont {S.}~\bibnamefont {Basak}}, \emph {et~al.},\
  }\bibfield  {title} {\bibinfo {title} {Second harmonic generation as a probe
  of broken inversion symmetry in weyl semimetals},\ }\href
  {https://doi.org/10.1038/s41567-018-0187-0} {\bibfield  {journal} {\bibinfo
  {journal} {Nature Physics}\ }\textbf {\bibinfo {volume} {14}},\ \bibinfo
  {pages} {900} (\bibinfo {year} {2018})}\BibitemShut {NoStop}%
\bibitem [{\citenamefont {Sipe}\ and\ \citenamefont {Shkrebtii}(2000)}]{sipe}%
  \BibitemOpen
  \bibfield  {author} {\bibinfo {author} {\bibfnamefont {J.~E.}\ \bibnamefont
  {Sipe}}\ and\ \bibinfo {author} {\bibfnamefont {A.~I.}\ \bibnamefont
  {Shkrebtii}},\ }\bibfield  {title} {\bibinfo {title} {Second-order optical
  response in semiconductors},\ }\href
  {https://doi.org/10.1103/PhysRevB.61.5337} {\bibfield  {journal} {\bibinfo
  {journal} {Phys. Rev. B}\ }\textbf {\bibinfo {volume} {61}},\ \bibinfo
  {pages} {5337} (\bibinfo {year} {2000})}\BibitemShut {NoStop}%
\bibitem [{\citenamefont {Khurgin}(1995)}]{khurgin}%
  \BibitemOpen
  \bibfield  {author} {\bibinfo {author} {\bibfnamefont {J.~B.}\ \bibnamefont
  {Khurgin}},\ }\bibfield  {title} {\bibinfo {title} {Current induced second
  harmonic generation in semiconductors},\ }\href
  {https://doi.org/10.1063/1.114978} {\bibfield  {journal} {\bibinfo  {journal}
  {Appl. Phys. Lett.}\ }\textbf {\bibinfo {volume} {67}},\ \bibinfo {pages}
  {1113} (\bibinfo {year} {1995})}\BibitemShut {NoStop}%
\bibitem [{\citenamefont {Bhalla}\ \emph {et~al.}(2022)\citenamefont {Bhalla},
  \citenamefont {Das}, \citenamefont {Culcer},\ and\ \citenamefont
  {Agarwal}}]{Bhalla2022}%
  \BibitemOpen
  \bibfield  {author} {\bibinfo {author} {\bibfnamefont {P.}~\bibnamefont
  {Bhalla}}, \bibinfo {author} {\bibfnamefont {K.}~\bibnamefont {Das}},
  \bibinfo {author} {\bibfnamefont {D.}~\bibnamefont {Culcer}},\ and\ \bibinfo
  {author} {\bibfnamefont {A.}~\bibnamefont {Agarwal}},\ }\bibfield  {title}
  {\bibinfo {title} {Resonant second-harmonic generation as a probe of quantum
  geometry},\ }\href {https://doi.org/10.1103/PhysRevLett.129.227401}
  {\bibfield  {journal} {\bibinfo  {journal} {Phys. Rev. Lett.}\ }\textbf
  {\bibinfo {volume} {129}},\ \bibinfo {pages} {227401} (\bibinfo {year}
  {2022})}\BibitemShut {NoStop}%
\bibitem [{\citenamefont {Singh}\ \emph {et~al.}(2017)\citenamefont {Singh},
  \citenamefont {Bolotin}, \citenamefont {Ghosh},\ and\ \citenamefont
  {Agarwal}}]{PhysRevB.95.155421}%
  \BibitemOpen
  \bibfield  {author} {\bibinfo {author} {\bibfnamefont {A.}~\bibnamefont
  {Singh}}, \bibinfo {author} {\bibfnamefont {K.~I.}\ \bibnamefont {Bolotin}},
  \bibinfo {author} {\bibfnamefont {S.}~\bibnamefont {Ghosh}},\ and\ \bibinfo
  {author} {\bibfnamefont {A.}~\bibnamefont {Agarwal}},\ }\bibfield  {title}
  {\bibinfo {title} {Nonlinear optical conductivity of a generic two-band
  system with application to doped and gapped graphene},\ }\href
  {https://doi.org/10.1103/PhysRevB.95.155421} {\bibfield  {journal} {\bibinfo
  {journal} {Phys. Rev. B}\ }\textbf {\bibinfo {volume} {95}},\ \bibinfo
  {pages} {155421} (\bibinfo {year} {2017})}\BibitemShut {NoStop}%
\bibitem [{\citenamefont {Cheng}\ \emph {et~al.}(2014)\citenamefont {Cheng},
  \citenamefont {Vermeulen},\ and\ \citenamefont {Sipe}}]{Cheng2014}%
  \BibitemOpen
  \bibfield  {author} {\bibinfo {author} {\bibfnamefont {J.-H.}\ \bibnamefont
  {Cheng}}, \bibinfo {author} {\bibfnamefont {N.}~\bibnamefont {Vermeulen}},\
  and\ \bibinfo {author} {\bibfnamefont {J.~E.}\ \bibnamefont {Sipe}},\
  }\bibfield  {title} {\bibinfo {title} {Third-order nonlinear optical response
  of graphene},\ }\href {https://doi.org/10.1088/1367-2630/16/5/053014}
  {\bibfield  {journal} {\bibinfo  {journal} {New J. Phys.}\ }\textbf {\bibinfo
  {volume} {16}},\ \bibinfo {pages} {053014} (\bibinfo {year}
  {2014})}\BibitemShut {NoStop}%
\bibitem [{\citenamefont {Hong}\ \emph {et~al.}(2013)\citenamefont {Hong},
  \citenamefont {Dadap}, \citenamefont {Petrone}, \citenamefont {Yeh},
  \citenamefont {Hone},\ and\ \citenamefont
  {Osgood~Jr.}}]{Hong2013THGGraphene}%
  \BibitemOpen
  \bibfield  {author} {\bibinfo {author} {\bibfnamefont {S.~Y.}\ \bibnamefont
  {Hong}}, \bibinfo {author} {\bibfnamefont {J.~I.}\ \bibnamefont {Dadap}},
  \bibinfo {author} {\bibfnamefont {N.}~\bibnamefont {Petrone}}, \bibinfo
  {author} {\bibfnamefont {P.-Y.}\ \bibnamefont {Yeh}}, \bibinfo {author}
  {\bibfnamefont {J.}~\bibnamefont {Hone}},\ and\ \bibinfo {author}
  {\bibfnamefont {R.~M.}\ \bibnamefont {Osgood~Jr.}},\ }\bibfield  {title}
  {\bibinfo {title} {Optical third-harmonic generation in graphene},\ }\href
  {https://doi.org/10.1103/PhysRevX.3.021014} {\bibfield  {journal} {\bibinfo
  {journal} {Phys. Rev. X}\ }\textbf {\bibinfo {volume} {3}},\ \bibinfo {pages}
  {021014} (\bibinfo {year} {2013})}\BibitemShut {NoStop}%
\bibitem [{\citenamefont {Pilch}\ \emph {et~al.}(2025)\citenamefont {Pilch},
  \citenamefont {Zhu}, \citenamefont {Kovalev}, \citenamefont {Dantas},
  \citenamefont {Bedoya-Pinto}, \citenamefont {Parkin},\ and\ \citenamefont
  {Wang}}]{Pilch2025THG}%
  \BibitemOpen
  \bibfield  {author} {\bibinfo {author} {\bibfnamefont {P.}~\bibnamefont
  {Pilch}}, \bibinfo {author} {\bibfnamefont {C.}~\bibnamefont {Zhu}}, \bibinfo
  {author} {\bibfnamefont {S.}~\bibnamefont {Kovalev}}, \bibinfo {author}
  {\bibfnamefont {R.~M.~A.}\ \bibnamefont {Dantas}}, \bibinfo {author}
  {\bibfnamefont {A.}~\bibnamefont {Bedoya-Pinto}}, \bibinfo {author}
  {\bibfnamefont {S.~S.~P.}\ \bibnamefont {Parkin}},\ and\ \bibinfo {author}
  {\bibfnamefont {Z.}~\bibnamefont {Wang}},\ }\bibfield  {title} {\bibinfo
  {title} {Terahertz third-harmonic generation of lightwave-driven weyl
  fermions far from equilibrium},\ }\href
  {https://doi.org/10.1021/acs.nanolett.5c04143} {\bibfield  {journal}
  {\bibinfo  {journal} {Nano Letters}\ }\textbf {\bibinfo {volume} {25}},\
  \bibinfo {pages} {16637} (\bibinfo {year} {2025})}\BibitemShut {NoStop}%
\bibitem [{\citenamefont {Mahan}(2000)}]{Mahan2000}%
  \BibitemOpen
  \bibfield  {author} {\bibinfo {author} {\bibfnamefont {G.~D.}\ \bibnamefont
  {Mahan}},\ }\href {https://doi.org/10.1007/978-1-4757-5714-9} {\emph
  {\bibinfo {title} {Many-Particle Physics}}},\ \bibinfo {edition} {3rd}\ ed.\
  (\bibinfo  {publisher} {Springer New York, NY},\ \bibinfo {year}
  {2000})\BibitemShut {NoStop}%
\bibitem [{\citenamefont {Eckardt}(2017)}]{Eckardt2017}%
  \BibitemOpen
  \bibfield  {author} {\bibinfo {author} {\bibfnamefont {A.}~\bibnamefont
  {Eckardt}},\ }\bibfield  {title} {\bibinfo {title} {Colloquium: Atomic
  quantum gases in periodically driven optical lattices},\ }\href
  {https://doi.org/10.1103/RevModPhys.89.011004} {\bibfield  {journal}
  {\bibinfo  {journal} {Rev. Mod. Phys.}\ }\textbf {\bibinfo {volume} {89}},\
  \bibinfo {pages} {011004} (\bibinfo {year} {2017})}\BibitemShut {NoStop}%
\bibitem [{\citenamefont {McIver}\ \emph {et~al.}(2020)\citenamefont {McIver},
  \citenamefont {Schulte}, \citenamefont {Stein}, \citenamefont {Matsuyama},
  \citenamefont {Jotzu}, \citenamefont {Meier},\ and\ \citenamefont
  {Cavalleri}}]{McIver2020}%
  \BibitemOpen
  \bibfield  {author} {\bibinfo {author} {\bibfnamefont {J.~W.}\ \bibnamefont
  {McIver}}, \bibinfo {author} {\bibfnamefont {B.}~\bibnamefont {Schulte}},
  \bibinfo {author} {\bibfnamefont {F.-U.}\ \bibnamefont {Stein}}, \bibinfo
  {author} {\bibfnamefont {T.}~\bibnamefont {Matsuyama}}, \bibinfo {author}
  {\bibfnamefont {G.}~\bibnamefont {Jotzu}}, \bibinfo {author} {\bibfnamefont
  {G.}~\bibnamefont {Meier}},\ and\ \bibinfo {author} {\bibfnamefont
  {A.}~\bibnamefont {Cavalleri}},\ }\bibfield  {title} {\bibinfo {title}
  {Light-induced anomalous hall effect in graphene},\ }\href
  {https://doi.org/https://doi.org/10.1038/s41567-019-0698-y} {\bibfield
  {journal} {\bibinfo  {journal} {Nat. Phys.}\ }\textbf {\bibinfo {volume}
  {16}},\ \bibinfo {pages} {38} (\bibinfo {year} {2020})}\BibitemShut {NoStop}%
\bibitem [{\citenamefont {Dabiri}\ \emph {et~al.}(2022)\citenamefont {Dabiri},
  \citenamefont {Cheraghchi},\ and\ \citenamefont {Sadeghi}}]{ali_Sadeghi}%
  \BibitemOpen
  \bibfield  {author} {\bibinfo {author} {\bibfnamefont {S.~S.}\ \bibnamefont
  {Dabiri}}, \bibinfo {author} {\bibfnamefont {H.}~\bibnamefont {Cheraghchi}},\
  and\ \bibinfo {author} {\bibfnamefont {A.}~\bibnamefont {Sadeghi}},\
  }\bibfield  {title} {\bibinfo {title} {Floquet states and optical
  conductivity of an irradiated two-dimensional topological insulator},\ }\href
  {https://doi.org/10.1103/PhysRevB.106.165423} {\bibfield  {journal} {\bibinfo
   {journal} {Phys. Rev. B}\ }\textbf {\bibinfo {volume} {106}},\ \bibinfo
  {pages} {165423} (\bibinfo {year} {2022})}\BibitemShut {NoStop}%
\bibitem [{\citenamefont {Parker}\ \emph {et~al.}(2019)\citenamefont {Parker},
  \citenamefont {Morimoto}, \citenamefont {Orenstein},\ and\ \citenamefont
  {Moore}}]{parker2019}%
  \BibitemOpen
  \bibfield  {author} {\bibinfo {author} {\bibfnamefont {D.~E.}\ \bibnamefont
  {Parker}}, \bibinfo {author} {\bibfnamefont {T.}~\bibnamefont {Morimoto}},
  \bibinfo {author} {\bibfnamefont {J.}~\bibnamefont {Orenstein}},\ and\
  \bibinfo {author} {\bibfnamefont {J.~E.}\ \bibnamefont {Moore}},\ }\bibfield
  {title} {\bibinfo {title} {Diagrammatic approach to nonlinear optical
  response with application to weyl semimetals},\ }\href
  {https://doi.org/10.1103/PhysRevB.99.045121} {\bibfield  {journal} {\bibinfo
  {journal} {Phys. Rev. B}\ }\textbf {\bibinfo {volume} {99}},\ \bibinfo
  {pages} {045121} (\bibinfo {year} {2019})}\BibitemShut {NoStop}%
\bibitem [{\citenamefont {Takasan}\ \emph {et~al.}(2021)\citenamefont
  {Takasan}, \citenamefont {Morimoto}, \citenamefont {Orenstein},\ and\
  \citenamefont {Moore}}]{moore2021}%
  \BibitemOpen
  \bibfield  {author} {\bibinfo {author} {\bibfnamefont {K.}~\bibnamefont
  {Takasan}}, \bibinfo {author} {\bibfnamefont {T.}~\bibnamefont {Morimoto}},
  \bibinfo {author} {\bibfnamefont {J.}~\bibnamefont {Orenstein}},\ and\
  \bibinfo {author} {\bibfnamefont {J.~E.}\ \bibnamefont {Moore}},\ }\bibfield
  {title} {\bibinfo {title} {Current-induced second harmonic generation in
  inversion-symmetric dirac and weyl semimetals},\ }\href
  {https://doi.org/10.1103/PhysRevB.104.L161202} {\bibfield  {journal}
  {\bibinfo  {journal} {Phys. Rev. B}\ }\textbf {\bibinfo {volume} {104}},\
  \bibinfo {pages} {L161202} (\bibinfo {year} {2021})}\BibitemShut {NoStop}%
\bibitem [{\citenamefont {Tavakol}\ and\ \citenamefont
  {Kim}(2023)}]{omidtavakol2023}%
  \BibitemOpen
  \bibfield  {author} {\bibinfo {author} {\bibfnamefont {O.}~\bibnamefont
  {Tavakol}}\ and\ \bibinfo {author} {\bibfnamefont {Y.~B.}\ \bibnamefont
  {Kim}},\ }\bibfield  {title} {\bibinfo {title} {Nonlinear optical responses
  in nodal line semimetals},\ }\href
  {https://doi.org/10.1103/PhysRevB.107.035114} {\bibfield  {journal} {\bibinfo
   {journal} {Phys. Rev. B}\ }\textbf {\bibinfo {volume} {107}},\ \bibinfo
  {pages} {035114} (\bibinfo {year} {2023})}\BibitemShut {NoStop}%
\bibitem [{\citenamefont {Giovannetti}\ \emph {et~al.}(2007)\citenamefont
  {Giovannetti}, \citenamefont {Khomyakov}, \citenamefont {Brocks},
  \citenamefont {Kelly},\ and\ \citenamefont {van~den
  Brink}}]{PhysRevB.76.073103}%
  \BibitemOpen
  \bibfield  {author} {\bibinfo {author} {\bibfnamefont {G.}~\bibnamefont
  {Giovannetti}}, \bibinfo {author} {\bibfnamefont {P.~A.}\ \bibnamefont
  {Khomyakov}}, \bibinfo {author} {\bibfnamefont {G.}~\bibnamefont {Brocks}},
  \bibinfo {author} {\bibfnamefont {P.~J.}\ \bibnamefont {Kelly}},\ and\
  \bibinfo {author} {\bibfnamefont {J.}~\bibnamefont {van~den Brink}},\
  }\bibfield  {title} {\bibinfo {title} {Substrate-induced band gap in graphene
  on hexagonal boron nitride: Ab initio density functional calculations},\
  }\href {https://doi.org/10.1103/PhysRevB.76.073103} {\bibfield  {journal}
  {\bibinfo  {journal} {Phys. Rev. B}\ }\textbf {\bibinfo {volume} {76}},\
  \bibinfo {pages} {073103} (\bibinfo {year} {2007})}\BibitemShut {NoStop}%
\bibitem [{\citenamefont {Chen}\ \emph {et~al.}(2025)\citenamefont {Chen},
  \citenamefont {Hou}, \citenamefont {Liang}, \citenamefont {Lu}, \citenamefont
  {Guo}, \citenamefont {Yan},\ and\ \citenamefont {Chang}}]{chen2025arxiv}%
  \BibitemOpen
  \bibfield  {author} {\bibinfo {author} {\bibfnamefont {X.}~\bibnamefont
  {Chen}}, \bibinfo {author} {\bibfnamefont {J.-T.}\ \bibnamefont {Hou}},
  \bibinfo {author} {\bibfnamefont {L.}~\bibnamefont {Liang}}, \bibinfo
  {author} {\bibfnamefont {J.}~\bibnamefont {Lu}}, \bibinfo {author}
  {\bibfnamefont {H.}~\bibnamefont {Guo}}, \bibinfo {author} {\bibfnamefont
  {C.-X.}\ \bibnamefont {Yan}},\ and\ \bibinfo {author} {\bibfnamefont {H.-R.}\
  \bibnamefont {Chang}},\ }\href {https://arxiv.org/abs/2510.06591} {\bibinfo
  {title} {Interband optical conductivity in two-dimensional semi-dirac bands
  tilting along the quadratic dispersion}} (\bibinfo {year} {2025}),\ \Eprint
  {https://arxiv.org/abs/2510.06591} {arXiv:2510.06591 [cond-mat.mes-hall]}
  \BibitemShut {NoStop}%
\end{thebibliography}%

\end{document}